# THERMODYNAMICS OF AMINE MIXTURES. SYSTEMS FORMED BY ALKYL-AMINE AND ETHER, OR *N,N*-DIALKYLAMIDE, OR ETHANENITRILE

Juan Antonio González, [(1)] Fernando Hevia,[(2)] Isaías García de la Fuente,[(1)] José Carlos Cobos[(1)], Karine Ballerat-Busserolles[(2)], Yohann Coulier[(2)], Jean-Yves Coxam[(2)]

[(1)]G.E.T.E.F., Departamento de Física Aplicada, Facultad de Ciencias, Universidad de Valladolid. Paseo de Belén, 7, 47011 Valladolid, Spain.

[(2)]Institut de Chimie de Clermont Ferrand, Thermodynamique et Interactions Moléculaires, UMR CNRS 6296, University Clermont Auvergne, Aubière, France

*e-mail: jagl@termo.uva.es; Tel: +34-983-423757

**Abstract**

Systems of the type linear primary or secondary amine + cyclohexane, or + polar (namely, linear or cyclic monoether, + 1,4-dioxane, + *N,N*-dialkylamide, or + ethanitrile) have been investigated using literature data, and by means of DISQUAC. Interaction parameters for the contacts amine/ether, amine/amide and amine/nitrile are provided. For a given contact, the QUAC interchange coefficients remain practically constant along each homologous series. A similar trend has been encountered in other many previous studies. DISQUAC correctly describes excess molar enthalpies, $H_m^E$, and vapour-liquid and solid-liquid equilibria of the studied mixtures and improves calculations on $H_m^E$ from the UNIFAC (Dortmund) model. The experimental data have been used to determine the enthalpy of the interactions between unlike molecules, which are stronger in systems with *N,N*-dialkylamides or ethanenitrile than in mixtures with ethers. On the other hand, it is shown that $H_m^E$ values of amine + $C_6H_{12}$ mixtures are closely related to the amine self-association, and that interactions between molecules of the polar compounds are determinant on $H_m^E$ results of the mixtures amine + fixed polar compound or of the systems fixed amine + polar compound (no linear monoether). Structural effects are relevant in the di-*n*-butylamine + linear ether systems. The application of the Flory model reveals that orientational effects are rather weak in the investigated solutions. This is in agreement with previous studies on this type of mixtures using the ERAS model.

## 1. Introduction

Linear primary and secondary amines (hereafter, alkyl-amines) are compounds with rather low dipole moments ( $\mu$ /D = 1.3 (1-butylamine); 1,2 (di-*n*-ethylamine) [1]; Table 1). In addition, they are self- associated substances [2-4] that form heterocomplexes when are mixed with other associated compounds [5-7]. As a consequence, the excess molar enthalpies, $H_{\mathrm{m}}^{\mathrm{E}}$, at 298.15 K and equimolar composition of 1-alkanol + linear primary or secondary amine mixtures are large and negative. Thus, at the mentioned conditions, $H_{\mathrm{m}}^{\mathrm{E}}$(methanol)/J.mol$^{-1}$ = $-3767$ [8] (1-butylamine); $-4581$ [9] (di-*n*-ethylamine). In this work, and as a continuation of our previous studies on amine + 2-alkanone mixtures (see, e.g. [10-14]), we investigate, using literature data, binary mixtures formed by alkyl-amines and several no self-associated compounds, that largely differ in their polarity. Particularly, the considered compounds are: cyclohexane (non-polar compound), or linear or cyclic monoethers or 1-4-dioxane, or *N,N*-dialkylamide o ethanenitrile (EtN). The latter compounds are strongly polar substances: $\mu$ /D = 3.68 (*N,N*-dimethylformamide, DMF) [15]; 3.53 (EtN) [16], and in pure liquid state present a significant local order, particularly DMF [17]. Regarding ethers, linear monooxalkanes are weakly polar substances ( $\mu$ = 1.2 D for di-*n*-ethylether [1]), while cyclic monoethers have higher dipolar moments (1.7 D for tetrahydrofuran [1], Table 1). A special case is 1,4-dioxane, which shows different conformations [18-20], and that in spite of its low $\mu$ value (0.4 D [1]), behaves as a polar compound [21]. Interestingly, it has been pointed out the difficulty of treating cyclic ethers using the classical UNIFAC model [22], and has been demonstrated, on the basis on principles of unchanging geometry and approximate group electroneutrality, that new groups are needed in order to provide a correct description of this type of systems [23,24]. In the present research, all the mixtures are treated in the framework of the DISQUAC model [25], and the corresponding interactions parameters for the contacts amine/X (X = c-$CH_2$, O, N-CO, CN) are reported. Binary systems containing linear primary [26] or secondary amines [27], or linear or cyclic ethers [28,29], or tertiary amide [30] or alkanenitriles [31] and alkanes have been already characterized in terms of DISQUAC. Except for the contacts amine/O (in linear ethers) and amine/N-CO in *N,N*-dimethylacetamide solutions, UNIFAC (Dortmund) interaction parameters for the remainder amine/X contacts mentioned above are available [32,33]. In a previous work [34], we have provided $H_{\mathrm{m}}^{\mathrm{E}}$ data for alkyl-amine + *N,N*-dialkylamide systems and applied the ERAS model [35] to these solutions. Alkyl-amine + linear or cyclic ethers mixtures [36,37], or the di-*n*-ethylamine + ethanenitrile system [38,39] have been also treated by means of ERAS, and the Kirkwood-Buff formalism [40,41] has been shortly applied to 1-propylamine, or 1-butylamine + di-*n*-butylether mixtures [42].

The functional groups considered in this work have great biological significance. For example, the disruption of amino acids releases amines, and proteins that are usually bound to DNA polymers contain several amine groups [43]. Histamine and dopamine are amines with the role of neurotransmitters [44,45]. Many hydrogen bonds are formed between nitriles and aminoacids such as it occurs in serine or arginine [46]. The study of systems including the amide functional group is necessary as a first step to a better knowledge of complex molecules of biological interest [47], and cyclic ethers have gained attraction in biotechnology [48,49]. It is also remarkable that many of the ionic liquids, which are very relevant in green chemistry, include amine groups [50].

Below, some of the compounds referred to are termed as follows: 1-butylamine (BA), 1-hexylamine (HxA), di-*n*-ethylamine (DEA), di-*n*-propylamine (DPA); di-*n*-butylamine (DBA); di-*n*-ethylether (DEE); di-*n*-propylether (DPE); di-*n*-butylether (DBE); tetrahydrofuran (THF); tetrahydropyran (THP); *N,N*-dimethylacetamide (DMA).

## 2. Survey of literature data

Literature data on vapour-liquid equilibria and $H_{\mathrm{m}}^{\mathrm{E}}$ for the systems considered along this research are shown in Tables 2-4 (see also Figures 1-6). The experimental excess molar Gibbs energies, $G_{\mathrm{m}}^{\mathrm{E}}$, were obtained by reducing the corresponding pressure-composition measurements by means of equations of the Redlich-Kister type with up to 3 coefficients. Non ideality of the vapour phase was taken into account using second virial coefficients determined according to the method of Hayden and O'Connell. Coordinates of the azeotropes, obtained from these correlations, are listed in Table 3. The direct $H_{\mathrm{m}}^{\mathrm{E}}$ measurements from the original papers have been also correlated with similar equations to those mentioned above using up to 5 coefficients. Tables 2 and 4 list, respectively, relative deviations for the pressure ($P$) and $H_{\mathrm{m}}^{\mathrm{E}}$ defined as:

$$\sigma_r(P) = \{\frac{1}{N}\sum\left[\frac{P_{\mathrm{exp}} - P_{\mathrm{calc}}}{P_{\mathrm{exp}}}\right]^2\}^{1/2} \quad (1)$$

and

$$dev(H_{\mathrm{m}}^{\mathrm{E}}) = \{\frac{1}{N}\sum\left[\frac{H_{\mathrm{m,exp}}^{\mathrm{E}} - H_{\mathrm{m,calc}}^{\mathrm{E}}}{H_{\mathrm{m,exp}}^{\mathrm{E}}(x_1 = 0.5)}\right]^2\}^{1/2} \quad (2)$$

where $N$ is the number of experimental points and $P_{\mathrm{calc}}$ and $H_{\mathrm{m,calc}}^{\mathrm{E}}$ are values obtained from the fitting equations. Inspection of Table 2 shows that $\sigma_r(P)$ values are within the expected experimental error. In contrast, $dev(H_{\mathrm{m}}^{\mathrm{E}})$ results are slightly large for some of the alkyl-amine

+ ether systems (Table 4). Nevertheless, the variation of $H_{\mathrm{m}}^{\mathrm{E}}(x_1 = 0.5)$ values with ether size is rather regular (Figure 7) allowing a consistent discussion of the experimental results. A similar statement is also valid for the remainder solutions (Figure 8). The reliability of the experimental data is important since in the present application the available database is limited and many different effects have to be considered. Thus, for alkyl-amine + DMF, or + DMA systems, they are: steric effects related to the increasing of the amine (amide) size in mixtures with a given amide (amine), or effects arising from the replacement of a linear primary amine by a secondary one in solutions with a given amide. Similar effects also exist in alkyl-amine + linear monoether mixtures and cyclization effects must be taken into account when cyclic ethers are present.

## 3. DISQUAC

This group contribution model is based on the rigid lattice theory developed by Guggenheim [51]. Some of its relevant features follow. (i) The total molecular volumes, $r_i$, surfaces, $q_i$, and the molecular surface fractions, $\alpha_{si}$, of the mixtures components are calculated additively on the basis of the group volumes $R_G$ and surfaces $Q_G$ recommended by Bondi [52]. At this end, the volume $R_{CH4}$ and surface $Q_{CH4}$ of methane are taken arbitrarily as volume and surface units [53]. The geometrical parameters for the groups referred to in this work are available in the literature [26-31,53] (ii) The partition function is factorized into two terms. Thus, the excess molar functions, Gibbs energies, and enthalpies, are calculated as the sum of two contributions. The dispersive (DIS) term represents the contribution from the dispersive forces; and the quasichemical (QUAC) term is due to the anisotropy of the field forces created by the solution molecules. In the case of $G_{\mathrm{m}}^{\mathrm{E}}$, a combinatorial term, $G_{\mathrm{m}}^{\mathrm{E,COMB}}$, given by the Flory-Huggins equation [53,54] must be included. Therefore,

$$G_{\mathrm{m}}^{\mathrm{E}} = G_{\mathrm{m}}^{\mathrm{E,DIS}} + G_{\mathrm{m}}^{\mathrm{E,QUAC}} + G_{\mathrm{m}}^{\mathrm{E,COMB}} \quad (3)$$

$$H_{\mathrm{m}}^{\mathrm{E}} = H_{\mathrm{m}}^{\mathrm{E,DIS}} + H_{\mathrm{m}}^{\mathrm{E,QUAC}} \quad (4)$$

(iii) The interaction parameters are assumed to be dependent on the molecular structure of the mixture components; (iv) The value $z = 4$ for the coordination number is used for all the polar contacts. This is an important shortcoming of DISQUAC and is partially removed considering structure dependent interaction parameters. (v) It is assumed that $V_{\mathrm{m}}^{\mathrm{E}}$ (excess molar volume) = 0.

The equations used to calculate the DIS and QUAC contributions to $G_{\mathrm{m}}^{\mathrm{E}}$ and $H_{\mathrm{m}}^{\mathrm{E}}$ within the theory are given elsewhere [5,25]. The temperature dependence of the interaction parameters is expressed in terms of the DIS and QUAC interchange coefficients [5,25], $C_{\mathrm{st,l}}^{\mathrm{DIS}}; C_{\mathrm{st,l}}^{\mathrm{QUAC}}$ where s $\neq$ t are two contact surfaces present in the mixture and $l$ = 1 (Gibbs energy; $C_{\mathrm{st},1}^{\mathrm{DIS/QUAC}} = g_{\mathrm{st}}^{\mathrm{DIS/QUAC}}(T_{\mathrm{o}}) / RT_{\mathrm{o}}$); $l$ = 2 (enthalpy, $C_{\mathrm{st},2}^{\mathrm{DIS/QUAC}} = h_{\mathrm{st}}^{\mathrm{DIS/QUAC}}(T_{\mathrm{o}}) / RT_{\mathrm{o}}$)), $l$ = 3 (heat capacity, $C_{\mathrm{st},3}^{\mathrm{DIS/QUAC}} = c_{\mathrm{pst}}^{\mathrm{DIS/QUAC}}(T_{\mathrm{o}}) / R$)). $T_{\mathrm{o}}$ = 298.15 K is the scaling temperature and *R,* the gas constant.

## 4. Adjustment of DISQUAC interaction parameters

In the framework of DISQUAC, alkyl-amine + cyclohexane, or + ether, or + *N,N*-dialkylamide, or + ethanenitrile mixtures are regarded as possessing the following four types of surface: (i) type a, aliphatic ($CH_3$, $CH_2$, H, in alkyl-amines, linear ethers, *N,N*-dialkylamide, or ethanenitrile); (ii) type c, cyclic (c-$CH_2$ in cyclohexane or cyclic oxaalkanes; (iii) type n, amine ($NH_2$ or NH in linear primary and secondary amines, respectively); (iv) type s (s = d, N-CO in *N,N*-dialkylamides; s = e, O in linear or cyclic ethers; and s = r, CN in ethanenitrile).

The general procedure applied in the estimation of the interaction parameters has been explained in detail elsewhere [5,25]. Final values of the fitted parameters in this work are collected in Table 5. Some important remarks are given below.

*4.1. Alkyl-amine + cyclohexane, or + linear ether, or + N,N*-dialkylamide, or + ethanenitrile

These solutions are built by three contacts: (a,n), (a,t), and (t,n), with t = c in cyclohexane systems, t = e, in mixtures with linear ethers, t = d in solutions including tertiary amides and t = r in ethanenitrile mixtures.T he interchange coefficients $C_{\mathrm{an},l}^{\mathrm{DIS}}$ and $C_{\mathrm{an},l}^{\mathrm{QUAC}}$ are known from the study of linear primary [26] or secondary [27] amine + alkane mixtures. Similarly, the DIS and QUAC interaction parameters for the (a,d) and (a,r) contacts are known from the corresponding DISQUAC treatments of *N,N*-dialkylamide [30] or alkanenitrile [31] + alkane mixtures. The $C_6H_{12}$ + *n*-alkane systems are characterized by dispersive interactions and therefore the (a,c) contacts are described by DIS parameters only [54].

*4.1.1. Alkyl-amine + cyclohexane*

Here, due to the interaction parameters for the (a,c) and (a,n) contacts are known, only those for (c,n) contacts must be determined. The corresponding adjustment was conducted assuming that $C_{\mathrm{an},l}^{\mathrm{QUAC}} = C_{\mathrm{cn},l}^{\mathrm{QUAC}}$ ($l$ =1,2). This is a rule widely used in the framework of DISQUAC, and it has been applied, e.g, when interaction parameters were determined for linear ether [28], or *n*-alkanone, [55], or linear organic carbonate [56], or chloroalkane [57], or *N,N*-dialkylamide [30], or alkanenitrile [31], or 1-alkanol [58] + cyclohexane mixtures once the corresponding *n*-alkane mixtures had been characterized using DISQUAC. The rule is very useful since it means that only the DIS parameters related to the c-$CH_2$/polar group contacts must be fitted, which is of particular interest when there is a lack of the experimental data required for the fitting of the parameters. On the other hand, here along the adjustment procedure, it was assumed that the dependence of the interaction parameters on the molecular structure of amines is the same to that encountered for the *n*-alkane systems:

(i) for mixtures including $CH_3(CH_2)_uNH_2$, the $C_{cn,l}^{DIS/QUAC}$ ($l$=1,2) coefficients remain unchanged from $u$ = 3 [26].

(ii) For systems with $CH_3(CH_2)_uNH(CH_2)_uCH_3$, the $C_{cn,l}^{DIS}$ ($l$ = 1,2) coefficients are independent of $u$ and $C_{cn,2}^{QUAC}$ is only different for $u$ = 1 [27].

*4.1.2. Alkyl-amine + N,N-dialkylamide*

For these solutions, only the interaction parameters for the (n,d) contacts must be determined since the interchange coefficients $C_{sn,l}^{DIS/QUAC}$ ($l$ =1,2; s = a,d) are known.

*4.1.3. Alkyl-amine + ethanenitrile*

Similarly, of the three contacts, (a,r), (a,n) and (n,r), which built these solutions only the interaction parameters for the latter remain unknown and must be fitted.

*4.2. Alkyl-amine + cyclic ether*

These systems are built by the following contacts: (a,c); (a,e); (a,n); (c,e), (c,n) and (e,n). The interaction parameters for (a,e) and (c,e) contacts are known from the study of cyclic ether + alkane mixtures [28,29]. Due to the $C_{cn,l}^{DIS/QUAC}$ coefficients have been determined above, only the DIS and QUAC parameters for the (n,e) contacts must be obtained.

## 5. Theoretical results

Comparisons between experimental data on VLE, $G_m^E$ and $H_m^E$ and theoretical results from the DISQUAC and UNIFAC models are shown along Tables 2-4. Tables 2 and 4 also contain, respectively, relative deviations for the pressure ($P$) and $H_m^E$ defined by equations (1) and (2), where $P_{calc}$ and $H_{m,calc}^E$ are now the corresponding DISQUAC values. Graphical comparisons between experimental and theoretical results can be seen in Figures 1-6 and Figures S1 and S2 of supplementary material

## 6. Discussion

Below we are referring to values of the excess functions at 298.15 K and equimolar composition.

Mixtures formed by alkyl-amine, or ether, or *N,N*-dialkylamide or ethanenitrile and alkane are characterized by showing positive $H_m^E$ values or miscibility gaps with more o less high upper critical solution temperatures (UCST). For example, $H_m^E$(heptane)/J.mol$^{-1}$ = 362 (DEE) [59]; 721 (DEA) [60]; 1192 (BA) [61]; 1338 (2-butanone) [62]; 1135 (2-pentanone) [62];

$H_m^E$(DMA + cyclohexane)/J.mol$^{-1}$ = 1687 [63]; and UCST(heptane)/K = 309.8 (DMA) [64]; 342.55 (DMF); [65]; 358 (EtN) [66]. This set of experimental data reveals that the thermodynamic properties of the mentioned solutions arise mainly from the breaking of interactions between like molecules along the mixing process.

Relative changes in intermolecular forces along a given homologous series of polar compounds or between homomorphic polar compounds can be examined by different ways. Firstly, the impact of polarity on bulk properties can be investigated by means of the effective dipole moment, $\bar{\mu}$, defined by [5,67,68]:

$$\bar{\mu} = \left[ \frac{\mu^2 N_A}{4\pi\varepsilon_0 V_m k_B T} \right]^{1/2} \tag{5}$$

Values of $\bar{\mu}$ for the pure polar compounds considered along the work are listed in Table 1. We note that while $\mu$ values change slightly along a homologous series, say 2-alkanones or linear monoethers, the corresponding $\bar{\mu}$ variation is sharper (Table 1). In fact, this magnitude decreases when the molecular size increases and, therefore, dipolar interactions become weaker at this condition. In addition, we must keep in mind that the potential energy of dipole-dipole interactions in a pure polar liquid is roughly proportional to $(-\mu^4 / V_m^2)$ [69]. For compounds with different polar groups (X = amine, O, CO, N-CO, CN), $\bar{\mu}$ changes in the sequence: 0.486 (DEE) < 0.500 (BA) < 0.996 (2-pentanone) < 1.12 (2-butanone) < 1.60 (DMF) < 1.86 (EtN) (Table 1). One can then conclude that dipolar interactions become stronger in the same order. Interestingly, BA and DEE have quite similar $V_m$ and $\bar{\mu}$ values while their $H_m^E$ results for heptane-containing mixtures largely differ. The same trend is observed for DEE and DEA, since the $\bar{\mu}$ value is lower for DEA (= 0.412) than for DEE, and the $H_m^E$ value of the DEA + heptane systems is higher. This merely underlines the importance of amine self-association when evaluating the $H_m^E$ values of their systems with alkanes. Secondly, the strength of the interactions between polar molecules can be also examined from the values of the partial excess molar enthalpy at infinite dilution of the first component, $H_{m1}^{E,\infty}$, corresponding to polar compound(1) + alkane(2) mixtures. Newly, we note that $H_{m1}^{E,\infty}$ decreases when the size of the polar compound is increased along a homologous series (Table 6) indicating that molecular interactions become weaker. For alkane mixtures differing in the polar group, $H_{m1}^{E,\infty}$/kJ.mol$^{-1}$ = 1.6 (DEE + heptane) [59]; 3.04 (DEA + heptane) [60]; 5.9 (BA + heptane) [61]; 6.35 (2-pentanone + heptane) [62]; 7.47 (2-butanone + heptane) [62]; 12.8 (DMA + cyclohexane); [63];

13.6 (DMF + cyclohexane) [70]; 15 (EtN + cyclohexane) [71]. Such variation is similar to that given above for $\overline{\mu}$. Finally, relative changes in intermolecular forces of homomorphic compounds can be also tentatively evaluated using the magnitude $\Delta\Delta H_{\text{vap}}$ defined by the equation [28,72,73]:

$$\Delta\Delta H_{\text{vap}} = \Delta H_{\text{vap}} \text{ (compound with a given polar group, X)} - \Delta H_{\text{vap}} \text{ (homomorphic hydrocarbon)} \quad (6)$$

In this equation, $\Delta H_{\text{vap}}$ is the standard enthalpy of vaporization at 298.15 K. Similar trends are encountered to those given above. Thus, $\Delta\Delta H_{\text{vap}}$ decreases along a homologous series when the size of the polar substance increases (Table 1), and $\Delta\Delta H_{\text{vap}}$ / kJ.mol$^{-1}$ = 0.62 (DEE) < 4.72 (DEA) < 9.09 (BA) < 12.03 (2-pentanone) < 13.7 (2-butanone) < 28.6 (DMA). In view of these results, if one considers molecules of similar size, molecular interactions between like molecules roughly become stronger in the sequence: linear ether < linear secondary amine < linear primary amine < 2-alkanone < DMA. It is clear that dipolar interactions are stronger in systems with DMF or EtN.

In mixtures with a given alkyl-amine, we note that $H_{\text{m}}^{\text{E}}$ (ether) < $H_{\text{m}}^{\text{E}}$ (alkane) (Fig. 7), while $H_{\text{m}}^{\text{E}}$ (alkyl-amine + 2-alkanone, or + DMF, or + DMA, or + EtN) < $H_{\text{m}}^{\text{E}}$ (polar compound + alkane) (Fig. 8). For example, $H_{\text{m}}^{\text{E}}$ (BA)/J.mol$^{-1}$ = 514 (DPE) [74] < 1192 (heptane) [61]; $H_{\text{m}}^{\text{E}}$ (2-butanone)/J.mol$^{-1}$ = 398 (DPA) [12] < 1338 (heptane) [62]; $H_{\text{m}}^{\text{E}}$ (DMA)/J.mol$^{-1}$ = 514 (DPA) [34] < 1687 (cyclohexane) [63]. As it is has been indicated, DMF, or EtN + heptane show miscibility gaps at 298.15 K, while, $H_{\text{m}}^{\text{E}}$ (BA)/J.mol$^{-1}$ = 564 (EtN, $T$ = 303.15 K) [75]; 386 (DMF) [34]. These relative variations can be ascribed to the creation of amine-compound interactions upon mixing.

The enthalpy of the interactions between alkyl-amines and a polar compound containing a functional group X (= O; CO; N-CO; CN) (termed as $\Delta H_{\text{N-X}}^{\infty}$) can be estimated as follows. Neglecting structural effects, it is possible to assume that $H_{\text{m}}^{\text{E}}$ is the result of three contributions [67,76]. Two of them, $\Delta H_{\text{N-N}}, \Delta H_{\text{X-X}}$, are positive, and come, respectively, from the disruption of amine-amine and X-X interactions upon the mixing. There is a negative third contribution, $\Delta H_{\text{N-X}}$, due to the new N---X interactions created along the mixing process. Therefore [77-80]:

$$H_{\text{m}}^{\text{E}} = \Delta H_{\text{N-N}} + \Delta H_{\text{X-X}} + \Delta H_{\text{N-X}} \quad (7)$$

For the studied systems, their positive $H_{\mathrm{m}}^{\mathrm{E}}$ values reveal that the contributions $\Delta H_{\text{N-N}}$ and $\Delta H_{\text{X-X}}$ are dominant over the $\Delta H_{\text{N-X}}$ term. Values of $\Delta H_{\text{N-X}}^{\infty}$ can be estimated assuming that equation (7) is held at $x_1 \rightarrow 0$ [77,79], and replacing $\Delta H_{\text{N-N}}$ and $\Delta H_{\text{X-X}}$ by $H_{\mathrm{m1}}^{\mathrm{E},\infty}$ of alkyl-amine(1) or polar compound(1) + heptane(2) systems. In such a case,

$$\Delta H_{\text{N-X}}^{\infty} = H_{\mathrm{m1}}^{\mathrm{E},\infty}(\text{alkyl}-\text{amine(1)} + \text{polar compound(2)})$$
$$-H_{\mathrm{m1}}^{\mathrm{E},\infty}(\text{alkyl}-\text{amine(1)} + \text{heptane(2)}) - H_{\mathrm{m1}}^{\mathrm{E},\infty}(\text{polar compound(1)} + \text{heptane(2)}) \quad (8)$$

We have applied this method to estimate the enthalpy of the amine-2-alkanone interactions [12,13], or of 1-alkanol-X interactions (e.g., X = O; CO, NO2; CN) [77,81-84] and now is used to evaluate the same magnitude for amine-X interactions (X = O; N-CO; CN) (Table 6). From inspection of Table 6 some statements can be provided. (i) For mixtures involving a given amine, more negative $\Delta H_{\text{N-X}}^{\infty}$ values are encountered for the systems linear primary amine + DMF, or + DMA or for alkyl-amine + EtN, which reveals that interactions between unlike molecules are stronger than for the remainder solutions. When comparing interactions between a given amine with different cyclic ethers, such interactions become stronger when 1,4-dioxane participates, since more negative $\Delta H_{\text{N-O}}^{\infty}$ values are encountered. (ii) Along systems with a fixed amine, $\Delta H_{\text{N-X}}^{\infty}$ values increase in line with the compound size in systems with linear monoethers, and are nearly constant for X = N-CO, or CO [12]. In such a case, it seems that steric effects related to the increase of alkyl groups attached to the polar group are of minor importance. (iii) The replacement of a linear monoether by a cyclic monoether in mixtures with a given amine (cyclization effect) leads to a strengthening of the interactions between unlike molecules, i.e., $\Delta H_{\text{N-O}}^{\infty}$(linear ether) > $\Delta H_{\text{N-O}}^{\infty}$(cyclic ether). (iv) For systems with a fixed compound including any of the functional groups X = O (1,4-dioxane), CO, or N-CO, $\Delta H_{\text{N-X}}^{\infty}$ values increase (are less negative) in line with the amine size. For example, $\Delta H_{\text{N-CO}}^{\infty}$(2-propanone)/kJ.mol$^{-1}$ = −8.1 (DPA); −5.9 (DBA) [12]. This can be due to the amine group is more sterically hindered in longer amines. (v) For the same reason, interactions between unlike molecules become weaker when a linear primary amine is replaced by a linear secondary one in systems with a fixed compound.

### *6.1 The effect of increasing the number of C atoms in the amine*

In alkyl-amine + cyclohexane systems, $H_{\mathrm{m}}^{\mathrm{E}}$ decreases when the amine size increases (Table 5), while in mixtures containing 1,4-dioxane or 2-alkanone, or DMF or DMA, $H_{\mathrm{m}}^{\mathrm{E}}$ increases (Table 5, Figure 8). In the former case, $H_{\mathrm{m}}^{\mathrm{E}}$ results change in line with the amine self-association. In the latter case, the variation of $H_{\mathrm{m}}^{\mathrm{E}}$ arises from: (i) less negative $\Delta H_{\text{N-X}}^{\infty}$ contribution for systems with longer amines, and (ii) increased $\Delta H_{\text{X-X}}$ values for the mentioned solutions since the larger aliphatic surfaces of longer amines are better breakers of the interactions between compound molecules. In fact, $H_{\mathrm{m}}^{\mathrm{E}}$ values of 2-alkanone, or 1,4-dioxane + *n*-alkane mixtures increase in line with alkane size [30,55] (see, eg., Figure 8). Note that the UCST of *N,N*-dialkylamide + *n*-alkane systems also increases in line with the alkane size: UCST(DMF)/K = 342.55 (*n*-$C_7$) [65]; 385.15 (*n*-$C_{16}$) [85].

*6.2 The effect of increasing the compound size in mixtures with a given amine*

Values of $H_{\mathrm{m}}^{\mathrm{E}}$ for BA + linear monoether systems increase with the oxaalkane size (Figure 7). This is explained by the corresponding decrease of $\left|\Delta H_{\text{N-O}}^{\infty}\right|$ (Table 6), and by the increasing of the $\Delta H_{\text{N-N}}^{\infty}$ contribution. This variation is the same to that for BA + *n*-alkane mixtures (Figure 7). In contrast, $H_{\mathrm{m}}^{\mathrm{E}}$ values decrease along a homologous series with *N,N*-dialkylamides or 2-alkanones [12] (Figure 8). Note that $\Delta H_{\text{N-X}}^{\infty}$ values are more or less constant, while the $\Delta H_{\text{X-X}}^{\infty}$ values become less positive (Table 6). The latter effect is dominant.

*6.3 The replacement of a linear primary amine by a linear secondary one in systems with a given compound*

$H_{\mathrm{m}}^{\mathrm{E}}$ values increase in systems containing compounds with X = O (in 1,4-dioxane), CN (Table 4) or N-CO (Figure 8) when a linear primary amine is replaced by a linear secondary one. Since $\Delta H_{\text{N-X}}^{\infty}$ values become less negative (Table 6), the observed increase for $H_{\mathrm{m}}^{\mathrm{E}}$ can be ascribed to the increasing from this contribution is prevalent.

As conclusion, $H_{\mathrm{m}}^{\mathrm{E}}$ results for amine + cyclohexane systems change accordingly to the amine-self-association; while $H_{\mathrm{m}}^{\mathrm{E}}$ values of mixtures of the type amine + fixed polar compound (X = O (1,4-dioxane), N-CO, CO; CN) change in line with the dipolar interactions between compound molecules. The same occurs for fixed amine + polar compound systems, with the exception of solutions including linear monoethers. It is remarkable that 1-alkanol + polar compound mixtures show similar trends. Thus, for $CH_3(CH_2)_uOH$ + DPE systems, $H_{\mathrm{m}}^{\mathrm{E}}$ increases up to $u = 3$ ($H_{\mathrm{m}}^{\mathrm{E}}$ = 742 J.mol$^{-1}$ [86]) and then slowly decreases ($H_{\mathrm{m}}^{\mathrm{E}}$ = 658 J.mol$^{-1}$, $u = 9$ [86]). 1-Alkanol + heptane mixtures behave similarly [77]. One can conclude that the alcohol self-association is determinant in systems with DPE. In contrast, $H_{\mathrm{m}}^{\mathrm{E}}$/J.mol$^{-1}$ increases in line with $u$ for systems with 2,5,8-trioxanonane: 440 ($u = 0$), 1624 ($u = 6$) [87]; or with 2-butanone: 708 ($u = 0$) [88]; 1844 ($u = 9$) [86] or with EtN: 1086 ($u = 0$) [89], 3032 ($u = 9$) [90] or with DMF: −103 ($u = 0$); 1146 ($u = 5$) [91]. Interestingly, different variations are encountered for $H_{\mathrm{m}}^{\mathrm{E}}$ of mixtures involving a given 1-alkanol when the compound size increases. For example, if DPE is replaced by DBE, $H_m^E$ increases. Thus, $H_m^E$($u$ = 0)/J.mol$^{-1}$ = 609 (DPE) [92]; 800 (DBE) [87], which supports our previous statement about that alkanol self-association plays a decisive role in these systems [77]. The replacement of DMF by DMA or of EtN by butanenitrile leads to decreased $H_m^E$ values. In fact, the experimental results (in J.mol$^{-1}$) for systems with $u$ = 0 and DMA ( – 737) [93], or butanenitrile (979) [94] are lower than those given above for the mixtures with DMF or EtN [89,95]. The corresponding polar compound + $n$-alkane mixtures behave similarly [30,31], and one can conclude that compound-compound interactions are crucial in systems formed by 1-alkanol and a polar compound of large dipole moment. It is to be noted that $H_m^E$ values also decrease along the series 1-hexanol + 2-alkanone, while are nearly constant for methanol + 2-alkane systems [96].

*6.4 Physical interactions.*

The relevance of this type of interactions can be demonstrated on the basis of the following considerations. (i) It is known that mixtures can be classified according to their position in the $G_m^E$ vs. $H_m^E$ diagram [97-99]. Firstly, we underline some important features of this type of diagrams. (a) $G_m^E = H_m^E/2$ is the dividing line between positive and negative values of $C_{pm}^E$ (isobaric excess molar heat capacities). Systems below this line show negative $C_{pm}^E$ values. The line $G_m^E = H_m^E$ divides the diagram in two parts with a different sign for $TS_m^E$. (b) In the first quarter of the plot, non-associated mixtures (e.g., cyclohexane + $n$-alkane) are situated between the lines $G_m^E = H_m^E/3$ and $G_m^E = H_m^E/2$. Such solutions are characterized by $C_{pm}^E$ < 0 and $TS_m^E$ > 0. (c) Mixtures with self-associated compounds (1-alkanol + alkane) are encountered in the region well above from the line $G_m^E = H_m^E$. These solutions have $C_{pm}^E$ > 0 and $TS_m^E$ < 0. (d) In the region between the lines $G_m^E = H_m^E/2$ and $G_m^E = H_m^E$, we find mixtures characterized by dipolar interactions (alkanone, or alkanoate + alkane mixtures). (e) If solvation exists (1-alkanol + alkyl-amine), then the solutions are situated in the third quarter of the diagram. For BA mixtures, we have $G_m^E$/J.mol$^{-1}$ = 318 (DBE) [42]; 210 (1,4-dioxane) [100] and $H_m^E$/J.mol$^{-1}$ = 632 (DBE), 523 (1,4-dioxane) [74]. Therefore, $TS_m^E$/J.mol$^{-1}$ values are positive: 314 (DBE); 313 (1,4-dioxane). The system with 1,4-dioxane is situated between the lines $G_m^E = H_m^E/3$ and $G_m^E$ = $H_m^E/2$, and dispersive interactions are dominant. Dipolar interactions are determinant in the DBE solution since is situated over the line $G_m^E = H_m^E/2$. Interestingly, the $TS_m^E$ value of the DEA + EtN mixture is negative ( – 155 J.mol$^{-1}$; $G_m^E$/J.mol$^{-1}$ = 767 [101]; $H_m^E$/J.mol$^{-1}$ = 612 [39]) and the system is situated above but close to the mentioned line $G_{\mathrm{m}}^{\mathrm{E}} = H_{\mathrm{m}}^{\mathrm{E}}$. This suggests that self-association effects are not very important. It must be remarked that 1-octylamine, or 1-decylamine + propanenitrile, or + butanenitrile mixtures show miscibility gaps [102], which clearly indicates the existence of strong dipolar interactions in such solutions. (ii) For amine + *N,N*-dialkylamide systems, $C_{\mathrm{pm}}^{\mathrm{E}}$ values are moderately positive. For DBA mixtures, the values (in J.mol$^{-1}$.K$^{-1}$) are 5 and 4.5 for the systems with DMF and DMA, respectively (F. Hevia, personal communication). Large positive $C_{\mathrm{pm}}^{\mathrm{E}}$ results are encountered for mixtures where the thermodynamic properties arise from the self-association of one component. Thus, for the ethanol + heptane mixture, $C_{\mathrm{pm}}^{\mathrm{E}}$ = 11.7 J.mol$^{-1}$.K$^{-1}$ [103]. In contrast, systems characterized by dipolar interactions show much lower $C_{\mathrm{pm}}^{\mathrm{E}}$ values (4.4 J.mol$^{-1}$.K$^{-1}$ for the 1-propanol + 2,5,8-trioxanonane system [104]). In such framework, the $C_{\mathrm{pm}}^{\mathrm{E}}$ result for the DEA + EtN mixture, 11.9 J.mol$^{-1}$.K$^{-1}$ [105] should be taken with caution since moreover largely differs from the value obtained from the $H_{\mathrm{m}}^{\mathrm{E}}$ change with the temperature (Table 4). (iii) Application of the ERAS model to alkyl-amine + ether, or + *N,N*-dialkylamide, or to DEA + EtN mixtures shows that the equilibrium constants, $K_{\mathrm{AB}}$, related to the formation of linear chains of the type $A_{\mathrm{n}}(\text{amine}) + B$ € $A_{\mathrm{n}}B$ ($B$ = ether, *N,N*-dialkylamide, EtN), are rather low, indicating that solvation effects can be neglected [34,36-39]. (iv) In addition, the physical parameter of the ERAS model is large for mixtures with *N,N*-dialkylamide [34] or EtN [39], which underlines the relevance of physical interactions in these systems. (v) We have shown [83,106-108] that the role of orientational effects in liquid mixtures can be investigated by means of the Flory model [109]. Here, this method has been shortly applied to a number of the considered systems. Results collected in Table 7 reveal that deviations between experimental and theoretical $H_{\mathrm{m}}^{\mathrm{E}}$ are low and that, therefore, orientational effects are weak in those mixtures. Stronger orientational effects are encountered for the BA + linear ether mixtures. Calculations on Kirkwood-Buff integrals for BA + DBE are in agreement with this conclusion [42].

*6.5 Structural effects and excess molar internal energies at constant volume*

The existence of structural effects in amine + 2-alkanone, or + *N,N*-dialkylamide mixtures has been previously discussed and will not be repeated here [12,13,34]. We pay attention to amine + ether, or + EtN systems. As a rule, one can state that structural effects are typical of systems which show positive $H_{\mathrm{m}}^{\mathrm{E}}$ values linked to negative $V_{\mathrm{m}}^{\mathrm{E}}$ results [110]. This is the case of the mixtures: BA + EtN (− 0.213 cm$^3$.mol$^{-1}$; $T$ = 303.15 K [111]); DEA + EtN (− 0.213 cm$^3$.mol$^{-1}$ [38]); BA + DEE (− 0.231 cm$^3$.mol$^{-1}$ [36]), or DEA + 1,4-dioxane (− 0.093 cm$^3$.mol$^{-1}$ [37]). Structural effects in DBA + linear monoether mixtures are clearly due to the difference in size between the system compounds, since $V_{\mathrm{m}}^{\mathrm{E}}$ approaches to zero when the mentioned difference vanishes ( – 0.334 and 0.028 $cm^3.mol^{-1}$ for the systems with DEE, or DBE, respectively [112]). It is interesting to investigate the influence of structural effects on $H_{\mathrm{m}}^{\mathrm{E}}$ due to this magnitude is not entirely determined by interactional effects, but also to structural effects [67,76]. The former are more properly considered using $U_{\mathrm{Vm}}^{\mathrm{E}}$, the excess internal energy at constant volume. If terms of higher order in $V_{\mathrm{m}}^{\mathrm{E}}$ are neglected, $U_{\mathrm{Vm}}^{\mathrm{E}}$ can be written as [67,76]:

$$U_{\mathrm{Vm}}^{\mathrm{E}} = H_{\mathrm{m}}^{\mathrm{E}} - \frac{\alpha_{\mathrm{p}}}{\kappa_{\mathrm{T}}} T V_{\mathrm{m}}^{\mathrm{E}} \qquad (9)$$

where $\frac{\alpha_p}{\kappa_T} T V_m^{\mathrm{E}}$ is the equation of state (eos) contribution to $H_{\mathrm{m}}^{\mathrm{E}}$, and $\alpha_p$ is the isobaric thermal expansion coefficient of the mixture, and $\kappa_T$, its isothermal compressibility. In this work, $\alpha_p$ and $\kappa_T$ were determined assuming ideal behavior ($F = \varphi_1 F_1 + \varphi_2 F_2$; $F_i$ is the property of the pure compound *i* for these magnitudes and $\varphi_{\mathrm{i}}$ is the volume fraction). We have demonstrated that the main features are the same for $H_{\mathrm{m}}^{\mathrm{E}}$ and $U_{\mathrm{Vm}}^{\mathrm{E}}$ for alkyl-amine + 2-alkanone, or + *N,N*-dialkylamide mixtures [12,13,34]. However, some of the amine + ether systems are characterized by low positive $H_{\mathrm{m}}^{\mathrm{E}}$ values and the contribution of structural effects to this magnitude may be important. Particularly, this occurs for DBA + linear monoether mixtures. We have $H_{\mathrm{m}}^{\mathrm{E}}$/J.mol$^{-1}$ = 53 (DEE), 69 (DPE), 73 (DBE) [74] and $U_{\mathrm{Vm}}^{\mathrm{E}}$/J.mol$^{-1}$ = 144 (DEE), 97 (DPE), 74 (DBE). That is, $U_{\mathrm{Vm}}^{\mathrm{E}}$ decreases in line with the difference in size of the mixture components since, at this condition, ether-ether interactions become weaker.

*6.6 Comparison between models*

UNIFAC and DISQUAC provide similar results for alkyl-amine + $C_6H_{12}$ mixtures (Tables 2 and 4). Poor $H_{\mathrm{m}}^{\mathrm{E}}$ results are obtained from UNIFAC for the systems alkyl-amine + cyclic ether, or + DMF, or + EtN. It is clear that the interaction parameters for the contacts amine/X (X = O in cyclic ether, N-CO or CN) need further improvement as it is indicated in the original works where the UNIFAC interaction parameters are reported [32,33]. Regarding the DISQUAC model, we underline the following points. (i) Differences between experimental and theoretical results may be somewhat large for some systems including cyclic ethers, characterized by low $H_{\mathrm{m}}^{\mathrm{E}}$ values. This is due to, in such cases, DISQUAC results are calculated as the difference between two large positive and negative values. However, even in such cases, the main features of the $H_{\mathrm{m}}^{\mathrm{E}}$ curves are described (Figure S2, supplementary material). (ii) The VLE of the cyclohexylamine + DMF system can be correctly represented (Tables 2 and 3) with the interaction parameters of the mixture with HxA by merely changing $C_{\mathrm{nd},1}^{\mathrm{DIS}}$ coefficient (Table 5). That is, the mentioned parameters are useful to describe the temperature dependence of VLE. (iii) Similarly, SLE data (see Table 8, Figure S3, supplementary material) for the systems 1-hexylamine, or 1-octylamine + EtN can be correlated using the interchange coefficients of the systems with 1-butylamine with new $C_{\mathrm{nd},1}^{\mathrm{DIS}}$ coefficients: 1.25 for 1-hexylamine and 3.0 for 1-octylamine (Table 5). This rapid increase of the $C_{\mathrm{nd},1}^{\mathrm{DIS}}$ coefficient is needed because the $G_{\mathrm{m}}^{\mathrm{E,COMB}}$ term sharply decreases when the difference in size between the mixture components increases. Thus, at equimolar composition and 298.15 K, $G_{\mathrm{m}}^{\mathrm{E,COMB}}$ (EtN)/J.mol$^{-1}$ = – 133 (1-butylamine); – 285 (1-hexylamine); – 433 (1-octylamine). (iv) We note that VLE and $H_{\mathrm{m}}^{\mathrm{E}}$ data for the DEA + EtN mixture are well represented by DISQUAC (Tables 2-4). It must be mentioned that the model predicts $C_{\mathrm{pm}}^{\mathrm{E}}$ = 5.6 J.mol$^{-1}$.K$^{-1}$ a value far from the no sure experimental result (see above).

*6.7 The dependence of DISQUAC interaction parameters with the molecular structure*

When we have investigated a number of systems using DISQUAC, we have encountered that the $C_{\mathrm{st},l}^{\mathrm{QUAC}}$ (l = 1,2) coefficients can be kept essentially constant along many of the homologous series under consideration. This is the case of the mixtures: pyridine [113] or *N,N*-dialkylamide [30], or crown ether [114] + *n*-alkane or 1-alkanol + *n*-alkanone [96], or + *n*-alkanoate [115] or + linear organic carbonate [116]. Nevertheless, it must be mentioned that some deviations from this behavior can exist for first member of the homologous series as in 1-alkanol + pyridine [113] or + *N,N*-dialkylamide [30] systems. Here, similar trends are encountered for the studied mixtures. Thus, the $C_{\mathrm{tn},l}^{\mathrm{QUAC}}$ (l =1,2) coefficients remain constant for the contacts (c,n), (e,n) or (d,n) along the corresponding homologous series, with different QUAC parameters for solutions including short alkyl-amines in cyclohexane mixtures, or for alkyl-amine + DEE systems (Table 5). A different $C_{\mathrm{en},2}^{\mathrm{QUAC}}$ coefficient is required to take into account properly the cyclization effect in mixtures involving ethers and linear secondary amines. This merely underlines the difficulty in treating cyclic molecules in the framework of group contribution models (see Introduction). Nevertheless, it is to be noted that the QUAC parameters are independent of the secondary amine under consideration. On the other hand, it seems that the same $C_{\mathrm{nr},l}^{\mathrm{QUAC}}$ ( l =1,2) coefficients can be used for mixtures with BA or DEA. The utility of this, say, rule can be demonstrated as follows. The 1-octylamine(1) + propanenitrile(2) mixture is not miscible over a very wide range of concentration and the liquid-liquid equilibrium curve is very shifted to very low amine mole fractions (UCST = 300 K at $x_1$ = 0.097 [102]). This complex behavior can be roughly described using the interaction parameters for 1-octylamine + EtN system (Table 5) with $C_{\text{nr},1}^{\text{DIS}}$ = 1.5. Then, the coordinates of the critical point are $x_1$ = 0.319 and UCST = 301.4 K

## 7. Conclusions

Interactions between unlike molecules are stronger in amine systems with *N,N*-dialkylamides or EtN than in solutions with ethers. The $H_{\text{m}}^{\text{E}}$ values of amine + $C_6H_{12}$ mixtures are mainly determined by the amine self-association. Compound-compound interactions are determinant on $H_{\text{m}}^{\text{E}}$ results of the mixtures amine + fixed polar compound or of fixed amine + polar compound (no linear monoether). Structural effects should be taken into account in the DBA + linear ether systems. Results from the Flory model reveal that orientational effects are rather weak in the systems under study. DISQUAC interaction parameters for the (n,t) (t = c, d, r) contacts are provided. The $C_{\text{nt},l}^{\text{QUAC}}$ (l = 1,2) coefficients are practically independent of the molecular structure along each homologous series. The model provides a good representation of $H_{\text{m}}^{\text{E}}$, VLE and SLE for the studied systems.

## 8. Acknowledgements

JAG, IGF and JCC gratefully acknowledge the financial support received from the Consejería de Educación y Cultura de Castilla y León, under Project VA100G19 (Apoyo a GIR, BDNS: 425389).

## 9. References

[1] R.C. Reid, J.M. Prausnitz and B.E. Poling. The Properties of Gases and Liquids, 4th Ed. McGraw Hill, N.Y., (1987).

[2] S. Villa, J.A. González, I. García de la Fuente, N. Riesco, J.C. Cobos, J. Solution Chem. 31 (2002) 1019-1038.

[3] X. Zhu, X. Cui, W. Cai, X. Shao, Acta Chim. Sin. 76 (2018) 298-302.

[4] A.J. Treszczanowicz, T. Treszczanowicz, Z. Phys. Chem. 217 (2003) 689-705

[5] J.A. González, I. García de la Fuente, J.C. Cobos, Fluid Phase Equilib. 168 (2000) 31-58.

[6] U. Domańska, M. Marciniak, Ind. Eng. Chem. Res. 43 (1994) 7647-7656.

[7] K.V. Zaitseva, M.A. Varfolomeev, B.N. Solomonov, Thermochim. Acta 535 (2012) 8-16

[8] A. Heintz, D. Papaioannou, Thermochim. Acta 310 (1998) 69-76.

[9] K. Nakanishi, H. Touhara, N. Watanabe, Bull. Chem. Soc. Jpn. 43 (1970) 2671–2676.

[10] I. Alonso, V. Alonso, I. Mozo, I. García de la Fuente, J.A. González, J.C. Cobos. J. Chem. Eng. Data, 55 (2010) 2505-2511.

[11] I. Alonso, I. Mozo, I. García de la Fuente, J.A. González, J.C. Cobos, J. Chem. Eng. Data 56 (2011) 3236-3241.

[12] J.A. González, I. Alonso, I. García de la Fuente, J.C. Cobos, Fluid Phase Equilib. 343 (2013) 1-12.

[13] J.A. González, I. Alonso, I. García de la Fuente, J.C. Cobos, Fluid Phase Equililib. 356 (2013) 117-125.

[14] I. Alonso, J.A. González, I. García de la Fuente, J.C. Cobos, J. Chem. Thermodyn. 69 (2014) 6-11

[15] A.L. McClellan, Tables of Experimental Dipole Moments, Vols., 1,2,3, Rahara Enterprises, El Cerrito, US, (1974).

[16] J.A. Riddick, W.B. Bunger, T. K. Sakano, Organic Solvents. Physical Properties and Methods of Purification, Techniques of Chemistry, Volume II. John Wiley & Sons, A Weissberger, Ed., N.Y, (1986).

[17] W.L. Jorgensen, C.L. Swenson, J. Am. Chem. Soc. 107 (1985) 569-578.

[18] R. Barata-Morgado, M.L Sánchez, I.F. Galván, J.C. Corchado, M.E. Martín, A. Muñoz-Losa, M.A. Aguilar, Theo. Chem. Acc. 132 (2013) 1390-11.

[19] P.I. Nagy, G. Volgyl, K. Takacs-Novak, J. Phys. Chem. B 112 (2008) 2085-2094.

[20] D.M. Chapman, R.E Hester, J. Phys. Chem. A 101 (1997) 3382-3387.

[21] M. Khajehpour, J.F. Kauffman, J. Phys. Chem. A 105 (2001) 10316-10321.

[22] H.S. Wu, S.I. Sandler, AIChE J. 25 (1989) 168-172.

[23] H.S. Wu, S.I. Sandler, Ind. Eng. Chem. Res. 30 (1991) 881-889.

[24] H.S. Wu, S.I. Sandler, Ind. Eng. Chem. Res. 30 (1991) 889-897.

[25] J.A. González, I. García de la Fuente, J.C. Cobos. Correlation and prediction of excess molar enthalpies using DISQUAC in: E. Wilhelm, T.M. Letcher (Eds.), Enthalpy and Internal Energy: Liquids, Solutions and Vapours, Royal Society of Chemistry, Croydon 2017.

[26] I. Velasco, J. Fernández, S. Otín, H.V. Kehiaian, Fluid Phase Equilib. 69 (1991) 15-32.

[27] M.R. Tiné, H.V. Kehiaian, Fluid Phase Equilib. 32 (1987) 211-248.

[28] H.V. Kehiaian, M.R. Tiné, L. Lepori, E. Matteoli, B. Marongiu, Fluid Phase Equilib. 46 (1989) 131-177.

[29] H.V. Kehiaian, M.R. Tiné, Fluid Phase Equilib. 59 (1990) 233-245.

[30] J.A. González, J.C. Cobos, I. García de la Fuente, Fluid Phase Equilib. 224 (2004) 169-183.

[31] B. Marongiu, B. Pittau, S. Porcedda, Thermochim. Acta 221 (1993) 143-162.

[32] J. Gmehling, J. Li, M. Schiller. Ind. Eng. Chem. Res., 32 (1993) 178-193.

[33] J. Gmehling, J. Lohmann, A. Jakob, J.Li, R. Joh, Ind. Eng. Chem. Res. 37 (1998) 4876-4882.

[34] F. Hevia, K. Ballerat-Busserolles, Y. Coulier, J.-Y. Coxam, J. A. González, I. García de la Fuente, J. C. Cobos, Fluid Phase Equilib. 502 (2019) 11283-12.

[35] A. Heintz, Ber. Bunsenges. Phys. Chem. 89 (1985) 172-181.

[36] T.M. Letcher, A. Goldon, Fluid Phase Equilib. 114 (1996) 147-159.

[37] T.M. Letcher, P.U. Govender, U. Domanska, J. Chem. Eng. Data 44 (1999) 274-285.

[38] R.B. Torres, A.Z. Francesconi, Fluid Phase Equilib. 200 (2002) 317-328.

[39] R.F. Checoni, A.Z. Francesconi, J. Therm. Anal. Calorim. 80 (2005) 295-301.

[40] J.G. Kirkwood, F.P. Buff. J. Chem. Phys. 19 (1954) 774-777.

[41] A. Ben-Naim. J. Chem. Phys. 67 (1977) 4884-4889.

[42] L. Lepori, E. Matteoli, L. Bernazzani, N. Ceccanti, G. Conti, P. Gianni, V. Mollica, M.R. Tiné, Phys. Chem. Chem. Phys. 2 (2000) 4837-4842.

[43] D.L. Nelson, M.M. Cox, Lehninger Principles of Biochemistry, 3rd ed., Worth Publishing, New York, (2000).

[44] L. Fong Fong, T. Hasegawa, M. Fukuda, E. Nakata, T. Morii, Bioorg. Med. Chem. 19 (1994) 4473-4481.

[45] J.M. Sonner, R.S. Cantor, Annu. Rev. Biophys., 42 (1994) 143-167.

[46] F.F. Fleming, L. Yao, P.C. Ravikumar, L. Funk, B.C. Shook, J. Med. Chem. 25 (2010) 7902-7917.

[47] E.S. Eberhardt, R.T. Raines, J. Am. Chem. Soc. 116 (1994) 2149-2150.

[48] Z. Feng, J. Zhao, Y. Li, S. Xu, J. Zhou, J. Zhang, L. Deng. A. Dong, Biomater. Sci. 4 (2016) 1493-1502.

[49] M. Kralj, L. Tusek-Bozic, L. Frkanec, ChemMedChem. 3 (2008) 1478-1492.

[50] M. Götz, R. Reimert, S. Bajohr, H. Schnetzer, J. Wimberg, T.J.S. Schubert, Thermochim. Acta 600 (2015) 82-88.

[51] E.A. Guggenheim, Mixtures, Oxford University Press, Oxford, 1952.

[52] A. Bondi, Physical Properties of Molecular Crystals, Liquids and Glasses, Wiley, New York, 1968.

[53] H.V. Kehiaian, J.-P.E. Grolier, G.C. Benson, J. Chim. Phys., 75 (1978) 1031-1048.

[54] J.A. González, I. García de la Fuente, J.C. Cobos, C. Casanova, A. Ait-Kaci, Fluid Phase Equilib. 112 (1995) 63-87.

[55] H.V. Kehiaian, S. Porcedda, B. Marongiu, L. Lepori, E. Matteoli, Fluid Phase Equilib. 63 (1991) 231-257

[56] J.A. González, I. García de la Fuente, J.C. Cobos, C. Casanova, H.V. Kehiaian. Thermochim. Acta, 217 (1993) 57-69

[57] H.V. Kehiaian, B. Marongiu, Fluid Phase Equilib. 40 (1988) 23-78.

[58] J.A. González, I. García de la Fuente, J.C. Cobos, C. Casanova, J. Solution Chem. 23 (1994) 399-420.

[59] B. Luo, G.C. Benson, B.C.-Y Lu, J. Chem. Thermodyn. 20 (1988) 267-271.

[60] E. Matteoli, P. Gianni, L. Lepori, Fluid Phase Equilib. 306 (2011) 234-241.

[61] E. Matteoli, L. Lepori, A. Spanedda, Fluid Phase Equilib. 212 (2003) 41-52.

[62] O. Kiyohara, Y.P. Handa, G.C. Benson, J. Chem. Thermodyn. 11 (1979) 453-460.

[63] H. Ukibe, R. Tanaka, S. Murakami, R. Fujishiro, J. Chem. Thermodyn. 6 (1974) 201-206.

[64] A. Xuequin, Z. Haihong, J. Fuguo, S. Weiguo, J. Chem. Thermodyn. 28 (1996) 1221-1232.

[65] J. Lobos, I. Mozo, M. Fernández Regúlez, J.A. González, I. García de la Fuente, J.C. Cobos, J. Chem. Eng. Data 51 (2006) 623-627.

[66] I.A. McLure, A.T. Rodríguez, P.A. Ingham, J.F. Steele, Fluid Phase Equilib. 8 (1982) 271-284.

[67] J.S. Rowlinson, F.L. Swinton, Liquids and Liquid Mixtures, third ed., Butterworths, London, 1982.

[68] E. Wilhelm, A. Laínez, J.-P.E. Grolier, Fluid Phase Equilib. 49 (1989) 233-250.

[69] T. Kimura, K. Suzuki, S. Takagi, Fluid Phase Equilib. 136 (1997) 269-278.

[70] J.N. Spencer, S.K. Berger, C.R. Powell, B.D. Henning, G.G. Furman,W.M. Loffredo, E.M. Rydberg, R.A. Neubert, C.T. Shoop, D.N. Blanch, J. Phys. Chem. 85 (1981) 1236-1241.

[71] M. Trampe, C.A. Eckert, J. Chem. Eng. Data 36 (1991) 112-118.

[72] J.A. González, I. García de la Fuente, J.C. Cobos, Fluid Phase Equilib. 154 (1999) 11-31.

[73] J.A. González, I. Alonso, C. Alonso-Tristán, I. García de la Fuente, J.C. Cobos, J. Chem. Thermodyn. 56 (2013) 89-98.

[74] T.M. Letcher, A. Goldon, J. Chem. Eng. Data 41 (1996) 629-633.

[75] G. Pathak, S.D. Adyanthaya, K.R. Patil, S.D. Pradhan, Thermochim. Acta 236 (1994) 123-130.

[76] H. Kalali, F. Kohler, P. Svejda, Fluid Phase Equilib. 20 (1985) 75-80.

[77] J.A. González, I. Mozo, I. García de la Fuente, J.C. Cobos, N. Riesco, J. Chem. Thermodyn. 40 (2008) 1495-1508.

[78] T.M. Letcher, U.P. Govender, J. Chem. Eng. Data 40 (1995) 1097-1100.

[79] T.M. Letcher, B.C. Bricknell, J. Chem. Eng. Data 41 (1996) 166-169.

[80] E. Calvo, P. Brocos, A. Piñeiro, M. Pintos, A. Amigo, R. Bravo, A.H. Roux, G. Roux-Desgranges, J. Chem. Eng. Data, 44 (1999) 948-954.

[81] J.A. González, A. Mediavilla, I. García de la Fuente, J.C. Cobos, J. Chem. Thermodyn. 59 (2013) 195-208.

[82] J.A. González, A. Mediavilla, I. García de la Fuente, J.C. Cobos, C. Alonso-Tristán, N. Riesco, Ind. Eng. Chem. Res. 52 (2013) 10317-10328.

[83] J.A. González, I. García de la Fuente, J.C. Cobos, C. Alonso-Tristán, L.F. Sanz, Ind. Eng. Chem. Res. 54 (2015) 550-559.

[84] J.A. González, F. Hevia, L.F. Sanz, I. García de la Fuente, C. Alonso-Tristán, Fluid Phase Equilib. 471 (2018) 24-39.

[85] M. Rogalski, R. Stryjek, Bull. Acad. Pol. Sci., Ser. Sci. Chim. XXVIII (1980) 139-146.

[86]. J. Iñarrea, J. Valero, P. Pérez, M. Gracia, C. Gutiérrez Losa, J. Chem. Thermodyn., 20 (1988) 193-199

[87] M.A. Villamañán, C. Casanova, A.H. Roux, J.-P.E. Grolier, J. Chem. Thermodyn. 14 (1982) 251-258

[88] I. Nagata, Int. DATA Ser., Sel. Data Mixtures, Ser. A. 2 (1984) 81.

[89] I. Nagata, K. Tamura, Fluid Phase Equilib. 24 (1985) 289-306.

[90] D.H. Lanfredi Viola, A.Z. Francesconi, J. Chem. Thermodyn. 47 (2012) 28-32.

[91] J.P. Chao, M. Dai, Y.X. Wang, J. Chem. Thermodyn. 21 (1989) 1169-1175.

[92] R. Garriga, F. Sánchez, P. Pérez, M. Gracia, J. Chem. Thermodyn. 29 (1997) 649-659.

[93] M. Oba, S. Murakami, R. Fujishiro, J. Chem. Thermodyn. 9 (1977) 407-414.

[94] R. Garriga, J. Ilarza, P. Pérez, M. Gracia, J. Chem. Thermodyn. 28 (1996) 233-243.

[95] J.A. González, F. Hevia, A. Cobos, I. García de la Fuente, C. Alonso-Tristán, Thermochim. Acta 605 (2015) 121-129.

[96] J.A. González, Can. J. Chem., 75 (1997) 1412-1423.

[97] F. Kohler, J. Gaube, Pol. J. Chem. 54 (1980) 1987-1993.

[98] R. Fuchs, L. Krenzer, J. Gaube, Ber. Bunsenges. Phys. Chem. 88 (1984) 642-649.

[99] K. P. Shukla, A.A. Chialvo, J.M. Haile, Ind. Eng. Chem. Res. 27 (1988) 664-671.

[100] I.L. Acevedo, M.A. Postigo, M. Katz J. Solution Chem. 17 (1988) 977-986.

[101] R. Srivastava, B.D. Smith, J. Chem. Eng. Data 30 (1985) 308-313.

[102] U. Domanska, M. Marciniak, J. Chem. Thermodyn. 39 (2007) 247-253.

[103] R. Tanaka, S. Toyama, S. Murakami, J. Chem. Thermodyn. 18 (1986) 63-73.

[104] M.A. Villamañán, C. Casanova, G. Roux-Desgranges, J.-P.E. Grolier, Thermochim. Acta, 52 (1982) 279-283.

[105] R.F. Checoni, A.Z. Francesconi, J. Solution Chem. 36 (2007) 913-922.

[106] J.A. González, N. Riesco, I. Mozo, I. García de la Fuente, J.C. Cobos, Ind. Eng. Chem. Res. 48 (2009) 7417-7429.

[107] J.A. González, Ind. Eng. Chem. Res. 49 (2010) 9511-9524.

[108] J.A. González, F. Hevia, C. Alonso-Tristán, I. García de la Fuente, J.C. Cobos, Fluid Phase Equilib. 449 (2017) 91-101.

[109] P.J. Flory, J. Am. Chem. Soc. 87 (1965) 1833-1838.

[110] L. Lepori, P. Gianni, E. Matteoli, J. Solution Chem. 42 (2013) 1263-1304.

[111] C.R. de Schaefer, F. Davolio, M. Katz, J. Solution Chem. 19 (1990) 289-299.

[112] T.M. Letcher, U. Domanska, P. Govender, J. Chem. Thermodyn. 26 (1994) 1019-1023.

[113] J.A. González, I. Mozo, I. García de la Fuente, J.C. Cobos, Thermochim. Acta, 441 (2006) 53-68.

[114] U. Domanska, J.A. González, Fluid Phase Equilib. 205 (2003) 317-338.

[115] J.A. González, I. Mozo, I. García de la Fuente, J.C. Cobos, Phys. Chem. Liq. 43 (2005) 175-194.

[116] J.A. González, M. Szurgocinska, U. Domanska, Fluid Phase Equilib. 200 (2002) 349-374.

[117] J. Canosa, A. Rodríguez, J. Tojo, Fluid Phase Equilib.156 (1999) 57-71.

[118] V. Majer, V. Svoboda, Enthalpies of Vaporization of Organic Compounds. Blackwell, Oxford, 1985.

[119] R. Garriga, S. Martínez, P. Pérez, M. Gracia, Fluid Phase Equilib. 147 (1998) 195-206.

[120] I. Mozo, I. García de la Fuente, J.A. González, J.C. Cobos, J. Mol. Liq. 129 (2006) 155-163.

[121] M. Keller, S. Schnabel, A. Heintz, Fluid Phase Equilib. 110 (1995) 231-265.

[122] D.R. Lide, CRC Handbook of Chemistry and Physics, 90th Edition, CRC Press/Taylor and Francis, Boca Raton, FL, 2010.

[123] R.F. Hudson, I. Stelzer, J. Chem. Soc. (B) (1966) 775-778.

[124] G.S. Shealy, S.I. Sandler, J. Chem. Eng. Data 30 (1985) 455-459.

[125] K. Engelmann, H.J. Bittrich, Wiss Z. Tech. Hoschsch. Chem. Lenna-Merseburg 11 (1969) 213 (cf. [142]).

[126] J. Fernández, I. Velasco, S. Otín. J. Chem. Thermodyn. 21 (1989) 419-422.

[127] A. Schumichen, PhD. Thesis (Leipzig), 1978 (cf. [143])

[128] I.L. Acevedo, G.C. Pedrosa, M. Katz, J. Chem. Eng. Data 41 (1996) 391-393.

[129] T.M. Letcher, U. Domanska, P. Govender, J. Chem. Thermodyn. 26 (1994) 681-689.

[130] F. Kimura, P. J. D'Arcy, G.C. Benson, J. Chem. Thermodyn. 15 (1983) 511-516.

[131] G.C. Benson, B. Luo, B.C.-Y. Lu, Can. J. Chem. 66 (1988) 531-534.

[132] A. Inglese, E. Wilhelm, J.-P. E. Grolier, H.V. Kehiaian, J. Chem. Thermodyn. 12 (1980) 217-222.

[133] S. Murakami, R. Fujishiro, Bull Chem. Soc Jpn. 39 (1966) 720-725.

[134] T.M. Letcher, J.W. Bayles, J. Chem. Eng. Data 16 (1971) 266-271.

[135] R. Pfestorf, D. Kuchenbecker, K. Quitzsch, Z. Phys. Chem. (Leipzig) 263 (1982) 233-240.

[136] M. Kern, L. Abello, D. Caceres, G. and Pannetier, Bull. Chim. Soc. Fr. 11 (1970) 3849-3856.

[137] H.-J. Bittrich, Wiss. Zeitschr. Techn. Hochschule Chemie Leuna-Merseburg, 8 (1966) 127-135.

[138] M.I. Paz-Andrade, F. Sarmiento, Int. DATA Ser., Sel. Data Mixtures, 2 (1984) 110-114.

[139] Y. Akamatsu, H. Ogawa, S. Murakami, Thermochim. Acta 113 (1987) 141-150.

[140] J.L. Legido, R. Bravo, M.I. Paz Andrade, L. Romani, F. Sarmiento, J. Ortega, J. Chem. Thermodyn. 18 (1986) 21-26.

[141] O. Urdaneta, Y.P. Handa, G.C. Benson, J. Chem. Thermodyn. 11 (1979) 857-860.

[142] J. Gmehling, U. Onken, J.R. Rarey, Vapor-Liquid Equilibrium Collection: Ethers (supplement 1) Vol . I, Part 4a, DECHEMA, Germany, 1982.

[143] C. Christensen, J. Gmehling, P. Rasmussen, U. Weidlich, Heat of Mixing Data Collection: Binary and Multicomponent Systems, Vol. III, Par 2, DECHEMA, Germany, 1984

TABLE 1

Physical properties of pure compounds at 298.15 K[a]

| Compound | $V_m$ / cm$^3$·mol$^{-1}$ | $\mu$ /D | $\bar{\mu}$ | $\Delta\Delta H_{vap}$ / kJ.mol$^{-1}$ |
|---|---|---|---|---|
| Di-*n*-ethylether (DEE) | 104.74 [117] | 1.3 [1] | 0.486 | 0.62 [118] |
| Di-*n*-propylether (DPE) | 137.68 [119] | 1.2 [1] | 0.396 | −0.88[118] |
| Di-*n*-butylether (DBE) | 170.45 [120] | 1.2 [1] | 0.352 | −1.43 [118] |
| Tetrahydrofuran (THF) | 81.75 [121] | 1.7 [1] | 0.720 | 3.44 [118] |
| Tetrahydropyran (THP) | 98.19 [122] | 1.6 [1] | 0.618 | 1.55 [118] |
| 1,4-dioxane | 85.71 [122] | 0.4 [1] | 0.165 | 5.54 [118] |
| Di-*n*-ethylamine (DEA) | 104.55 [60] | 1.1 [1] | 0.412 | 4.72 [118] |
| Di-n-propylamine (DPA) | 137.91 [60] | 1.0 [1] | 0.326 | 3.44 [118] |
| Di-*n*-butylamine (DBA) | 171.09 [60] | 1.1 [1] | 0.322 | 3.04 [118] |
| 1-butylamine (BA) | 99.88 [61] | 1.3 [1] | 0.498 | 9.09 [118] |
| 1-hexylamine (HxA) | 133.06 [61] | 1.3 [15] | 0.431 | 8.47 [118] |
| 2-propanone | 74.04 [16] | 2.88 [122] | 1.28 | 15.02 [118] |
| 2-butanone | 90.17 [16] | 2.78 [122] | 1.12 | 13.3 [118] |
| 2-heptanone | 140.76 [16] | 2.59 [122] | 0.836 | 10.6 [118] |
| *N,N*-dimethylformamide (DMF) | 77.44 [16] | 3.68 [15] | 1.60 | 30.6 [118] |
| *N,N*-dimethylacetamide (DMA) | 93.04 [16] | 3.71 [15] | 1.49 | 28.6 [118] |
| Ethanenitrile (EtN) | 52.87 [16] | 3.53 [16] | 1.86 | 23.6 [118] |

[a] $V_m$, molar volume; $\mu$, dipole moment; $\bar{\mu}$, effective dipole moment (equation 5); $\Delta\Delta H_{vap}$, differences between the standard enthalpy of vaporization at 298.15 for a compound with a given polar group and that of the corresponding homomorphic hydrocarbon (equation 6).

TABLE 2

Molar excess Gibbs energies, $G_m^E$, at equimolar composition and at temperature *T*, for alkyl-amine(1) + organic compound (2) mixtures.

| System[a] | $N^b$ | $T$/K | $G_m^E$/J·mol$^{-1}$ | | $\sigma_r(P)^c$ | | Ref. |
|---|---|---|---|---|---|---|---|
| | | | Exp[d]. | DQ[e] | Exp[d,f]. | DQ[e]/UNIF[g] | |
| DEA + $C_6H_{12}$ | 7 | 302.25 | 251 | 248 | 0.007 | 0.008/0.008 | 123 |
| | 6 | 314.25 | 201 | 230 | 0.004 | 0.012/0.007 | 123 |
| PA + DBE | 15 | 298.15 | 376 | 372 | 0.002 | 0.068/n.a | 42 |
| BA + DBE | 18 | 298.15 | 318 | 316 | 0.008 | 0.050/n.a | 42 |
| BA + 1,4-dioxane | 17 | 298.15 | 204 | 223 | 0.003 | 0.008/0.005 | 100 |
| Cyclohexyalmine + DMF | 16 | 373.15 | 651 | 672 | 0.002 | 0.013/n.a | 124 |
| | 15 | 393.15 | 655 | 670 | 0.003 | 0.008/n.a | 124 |
| DEA + 1,4-dioxane | 8 | 338.15 | 338 | 331 | 0.010 | 0.011/0.017 | 125 |
| | 8 | 343.15 | 333 | 320 | 0.010 | 0.012/0.021 | 125 |
| DEA + EtN | 13 | 298.00 | 767 | 777 | 0.001 | 0.008/0.018 | 101 |
| | 13 | 347.93 | 795 | 783 | 0.0008 | 0.007/0.004 | 101 |
| | 13 | 398.33 | 774 | 760 | 0.002 | 0.005/0.029 | 101 |

[a]for symbols, see Table 1; PA, 1-propylamine; [b]number of data points; [c]equation (1); [d]experimental value; [e]DISQUAC value calculated with interaction parameters from Table 5; [f]value obtained from the fitting of experimental results to Redlich-Kister expansions; [g]UNIFAC results using interaction parameters literature [32,33] (n.a., not available).

TABLE 3

Coordinates of azeotropic points: temperature ($T_{az}$), composition ($x_{1az}$) and pressure ($P_{az}$) for amine (1) + organic compound (2) mixtures.

| System[a] | $T_{az}$ /K | $x_{1az}$ | | $P_{az}$ /kPa | | Ref. |
|---|---|---|---|---|---|---|
| | | Exp.[b] | DQ.[b] | Exp.[b] | DQ.[c] | |
| DEA + EtN | 298 | 0.883 | 0.884 | 32.12 | 32.09 | 101 |
| | 398.33 | 0.828 | 0.837 | 702.9 | 699.9 | 101 |
| Cyclohexylamiane + DMF | 373.15 | 0.862 | 0.874 | 36.56 | 36.20 | 124 |
| | 393.15 | 0.862 | 0.858 | 68.85 | 68.49 | 124 |

[a] for symbols, see Table 1; [b]experimental value; [c]DISQUAC value calculated with interaction parameters from Table 5

TABLE 4

Molar excess enthalpies, $H_m^E$, at equimolar composition and at 298.15 K, for alkyl-amine(1) + organic compound (2) mixtures.

| System[a] | $N$[b] | $T$/K | $H_m^E$/J·mol$^{-1}$ | | $dev(H_m^E)$[c] | | Ref. |
|---|---|---|---|---|---|---|---|
| | | | Exp[d]. | DQ[e] | Exp[d,f]. | DQ[e]/UNIF[g] | |
| Linear primary amine + cyclohexane | | | | | | | |
| $CH_3(CH_2)_2NH_2 + C_6H_{12}$ | 15 | 303.15 | 1243 | 1267 | 0.005 | 0.015/0.045 | 126 |
| $CH_3(CH_2)_3NH_2 + C_6H_{12}$ | 24 | 298.15 | 1200 | 1201 | 0.004 | 0.009/0.032 | 127 |
| | 17 | 303.15 | 1131 | 1198 | 0.004 | 0.044/0.031 | 126 |
| $CH_3(CH_2)_5NH_2 + C_6H_{12}$ | 26 | 298.15 | 1042 | 974 | 0.005 | 0.063/0.061 | 127 |
| | 25 | | 987 | | 0.010 | 0.017/0.016 | 128 |
| | 18 | 303.15 | 983 | 965 | 0.004 | 0.013/0.010 | 126 |
| | 25 | 313.15 | 991 | 947 | 0.016 | 0.047/0.021 | 128 |
| $CH_3(CH_2)_7NH_2 + C_6H_{12}$ | 19 | 298.15 | 988 | 848 | 0.006 | 0.121/0.122 | 127 |
| | 20 | 303.15 | 885 | 835 | 0.004 | 0.041/0.025 | 126 |
| $CH_3(CH_2)_9NH_2 + C_6H_{12}$ | 22 | 298.15 | 938 | 785 | 0.003 | 0.139/0.153 | 127 |
| | 18 | 303.15 | 840 | 766 | 0.005 | 0.061/0.059 | 126 |
| $CH_3(CH_2)_{11}NH_2 + C_6H_{12}$ | 19 | 303.15 | 826 | 738 | 0.005 | 0.074/0.105 | 126 |
| $CH_3(CH_2)_{14}NH_2 + C_6H_{12}$ | 13 | 313.15 | 769 | 682 | 0.006 | 0.077/0.115 | 126 |
| Alkyl-amine + ether | | | | | | | |
| BA + DEE | 20 | 298.15 | 264 | 269 | 0.017 | 0.025/n.a | 74 |
| BA + DPE | 20 | 298.15 | 514 | 503 | 0.014 | 0.041/n.a | 74 |
| BA + DBE | 23 | 298.15 | 632 | 624 | 0.014 | 0.028/n.a | 74 |
| BA + THF | 13 | 298.15 | 154 | 155 | 0.025 | 0.041/0.853 | 74 |
| BA+ THP | 15 | 298.15 | 223 | 231 | 0.026 | 0.039/0.628 | 74 |
| BA + 1,4 dioxane | 14 | 298.15 | 523 | 537 | 0.013 | 0.055/0.808 | 74 |
| | | | 523 | | 0.010 | 0.073/0.792 | 128 |
| DEA + THF | 25 | 298.15 | 51 | 50 | 0.033 | 0.066/0.52 | 37 |
| DEA+ THP | 23 | 298.15 | 27 | 24 | 0.048 | 0.242/3.71 | 37 |
| DEA + 1,4-dioxane | 20 | 298.15 | 627 | 631 | 0.019 | 0.027/0.654 | 37 |
| DPA + THF | 17 | 298.15 | 59 | 55 | 0.019 | 0.305/1.27 | 37 |
| DPA + THP | 16 | 298.15 | 14 | 14 | 0.093 | 0.386/8.91 | 37 |
| DPA + 1,4-dioxane | 15 | 298.15 | 713 | 718 | 0.013 | 0.038/0.372 | 37 |
| DBA + DEE | 22 | 298.15 | 52 | 51 | 0.035 | 0.062/n.a | 129 |
| DBA + DPE | 22 | 298.15 | 69 | 68 | 0.022 | 0.064/n.a | 129 |
| DBA + DBE | 26 | 298.15 | 73 | 71 | 0.026 | 0.060/n.a | 129 |

TABLE 4 (continued)

| | | | | | | | |
|---|---|---|---|---|---|---|---|
| DBA + THF | 20 | 298.15 | 104 | 102 | 0.017 | 0.035/0.748 | 129 |
| DBA + THP | 21 | 298.15 | 63 | 60 | 0.017 | 0.054/2.57 | 129 |
| DBA + 1,4dioxane | 28 | 298.15 | 981 | 990 | 0.014 | 0.062/0.285 | 129 |
| Alkyl-amine + *N,N*-dialkylamide | | | | | | | |
| BA + DMF | 15 | 298.15 | 386 | 373 | 0.005 | 0.044/0.622 | 34 |
| HxA + DMF | 14 | 298.15 | 660 | 655 | 0.006 | 0.020/0.434 | 34 |
| DPA + DMF | 14 | 298.15 | 750 | 755 | 0.005 | 0.013/1.23 | 34 |
| DBA + DMF | 14 | 298.15 | 1102 | 1104 | 0.007 | 0.032/0.863 | 34 |
| BA + DMA | 17 | 298.15 | 208 | 218 | 0.012 | 0.053/n.a | 34 |
| HxA + DMA | 19 | 298.15 | 425 | 412 | 0.007 | 0.061/n.a | 34 |
| DPA + DMA | 15 | 298.15 | 509 | 520 | 0.007 | 0.033/n.a | 34 |
| DBA + DMA | 15 | 298.15 | 828 | 843 | 0.005 | 0.013/n.a | 34 |
| Alkyl-amine + ethanenitrile | | | | | | | |
| BA + EtN | 8 | 303.15 | 564 | 572 | 0.013 | 0.043/0.721 | 75 |
| DEA + EtN | 33 | 288.15 | 552 | 563 | 0.004 | 0.014/0.316 | 39 |
| | 33 | 298.15 | 612 | 621 | 0.004 | 0.021/0.232 | 39 |
| | 33 | 303.15 | 641 | 649 | 0.004 | 0.030/0.491 | 39 |

[a]for symbols, see Table 1; [b]number of data points; [c]equation (2); [d]experimental value; [e]DISQUAC value calculated with interaction parameters from Table 5; [f]value obtained from the fitting of experimental results to Redlich-Kister expansions; [g]UNIFAC results using interaction parameters literature [32,33] (n.a., not available)..

TABLE 5

Dispersive (DIS) and quasichemical (QUAC) interchange coefficients, $C^{\mathrm{DIS}}_{\mathrm{nt},l}$ and $C^{\mathrm{QUAC}}_{\mathrm{nt},l}$ ($l$ = 1, Gibbs energy; $l$ = 2, enthalpy; $l$ = 3, heat capacity) for (n,s) contacts[a] in alkyl-amine + organic compound mixtures

| System[b] | $C^{\mathrm{DIS}}_{\mathrm{nt},1}$ | $C^{\mathrm{DIS}}_{\mathrm{nt},2}$ | $C^{\mathrm{DIS}}_{\mathrm{nt},3}$ | $C^{\mathrm{QUAC}}_{\mathrm{nt},1}$ | $C^{\mathrm{QUAC}}_{\mathrm{nt},2}$ | $C^{\mathrm{QUAC}}_{\mathrm{nt},3}$ |
|---|---|---|---|---|---|---|
| Linear primary amine + cyclohexane (t = c-$CH_2$) | | | | | | |
| $CH_3(CH_2)_2NH_2$+ $C_6H_{12}$ | 1 | 1.9 | | 3.549 | 7.269 | |
| $CH_3(CH_2)_uNH_2$ + $C_6H_{12}$ ($u$ > 2) | 1.15 | 2.6 | | 3.51 | 7.20 | |
| Linear secondary amine + cyclohexane (t = c-$CH_2$) | | | | | | |
| $CH_3CH_2NHCH_2CH_3$ + $C_6H_{12}$ | 3.35 | 9.7 | | 3.105 | 6.78 | |
| $CH_3(CH_2)_uNH(CH_2)_uCH_3$ + $C_6H_{12}$ ($u$ =2,3) | 3.35 | 9.7 | | 3.105 | 5.48 | |
| Alkyl-amine + ether (t = O) | | | | | | |
| $CH_3(CH_2)_2NH_2$ + DBE | 8.5 | 3.7 | | 1 | −3 | |
| $CH_3(CH_2)_3NH_2$ + DEE | 6.5 | 11.8 | | 1 | −3.8 | |
| $CH_3(CH_2)_3NH_2$ + DPE | 6.5 | 8.5 | | 1 | −3. | |
| $CH_3(CH_2)_3NH_2$ + DBE | 6.5 | 2.9 | | 1 | −3. | |
| $CH_3(CH_2)_3NH_2$ + THF | 3.2 | 17 | | 1 | −3. | |
| $CH_3(CH_2)_3NH_2$ + THP | 3.2 | 9.8 | | 1 | −3. | |
| $CH_3(CH_2)_3NH_2$+ 1,4-dioxane | 3.2 | 15.4 | | 1 | −3. | |
| $CH_3CH_2NHCH_2CH_3$ + THF | 3.2 | 26 | | 1 | −7. | |
| $CH_3CH_2NHCH_2CH_3$ + THP | 3.2 | 13.6 | | 1 | −7. | |
| $CH_3CH_2NHCH_2CH_3$ + 1,4-dioxane | 3.2 | 17.4 | | 1 | −3. | |
| $CH_3(CH_2)_2NH(CH_2)_2CH_3$ + THF | 3.2 | 23.6 | | 1 | −7. | |
| $CH_3(CH_2)_2NH(CH_2)_2CH_3$ + THP | 3.2 | 11.2 | | 1 | −7. | |
| $CH_3(CH_2)_2NH(CH_2)_2CH_3$ + 1,4-dioxane | 3.2 | 12.6 | | 1 | −3. | |
| $CH_3(CH_2)_3NH(CH_2)_3CH_3$ + DEE | 6.5 | 8.9 | | 1 | −3.8 | |
| $CH_3(CH_2)_3NH(CH_2)_3CH_3$ + DPE | 6.5 | 8.9 | | 1 | −3. | |
| $CH_3(CH_2)_3NH(CH_2)_3CH_3$ + DBE | 6.5 | 5.1 | | 1 | −3. | |
| $CH_3(CH_2)_3NH(CH_2)_3CH_3$ + THF | 3.2 | 21.6 | | 1 | −7. | |
| $CH_3(CH_2)_3NH(CH_2)_3CH_3$ + THP | 3.2 | 9.2 | | 1 | −7. | |
| $CH_3(CH_2)_3NH(CH_2)_3CH_3$ + 1,4-dioxane | 3.2 | 14.9 | | 1 | −3. | |
| Alkyl-amine + *N,N*-dialkylamide (t = d) | | | | | | |
| $CH_3(CH_2)_uNH_2$ + DMF ($u$ = 3,5) | 0.5 | −2.8 | | 1 | 2 | |
| $CH_3(CH_2)_uNH_2$ + DMA ($u$ = 3,5) | 0.5 | −3.2 | | 1 | 2 | |
| $C_6H_{11}NH_2$ + DMF | 4.05 | −2.8 | | 1 | 2 | |

Table 5 (continued)

| | | | | | | |
|---|---|---|---|---|---|---|
| $CH_3(CH_2)_2NH(CH_2)_2CH_3$ + DMF | 0.5 | 1 | | 1 | − 0.8 | |
| $CH_3(CH_2)_3NH(CH_2)_3CH_3$ + DMF | 0.5 | 2.6 | | 1 | − 0.8 | |
| $CH_3(CH_2)_2NH(CH_2)_2CH_2$ + DMA | 0.5 | 2 | | 1 | − 0.8 | |
| $CH_3(CH_2)_3NH(CH_2)_3CH_3$ + DMA | 0.5 | 4.5 | | 1 | − 0.8 | |
| Alkyl-amine + ethanenitrile (t = r) | | | | | | |
| $CH_3(CH_2)_3NH_2$ + EtN | 1.25[c] | 5.1 | | 0.7 | − 1 | |
| $CH_3(CH_2)_5NH_2$ + EtN | 1.25 | 5.1 | | 0.7 | − 1 | |
| $CH_3(CH_2)_7NH_2$ + EtN | 3.0 | 5.1 | | 0.7 | − 1 | |
| $CH_3CH_2NHCH_2CH_3$ + EtN | 0.55 | 9.9 | − 8 | 0.7 | − 1 | 6 |

[a]type c, c-$CH_2$ in cyclohexane or cyclic ethers; type e, O in linear or cyclic ethers; type d, N-CO in *N,N*-dialkylamides: type n, $NH_2$ linear primary amines, or NH in linear secondary amines, type r, CN in ethanenitrile; [b]for symbols, see Table 1; [c]estimated value extrapolating $G_m^E$ values for the remainder alkyl-amine + EtN considered.

TABLE 6

Partial excess molar enthalpies, $H_1^{E,\infty}$, at $T$ = 298.15 K for solute(1) + organic compound(2) mixtures, and enthalpy of the amine-compound interaction, $\Delta H_{\text{N-X}}^{\infty}$ (eq. 8), at $T$ = 298.15 K, for Alkyl-amine(1) + organic compound(2) systems (X = O; N-CO, CN).

| System[a] | $H_{m,1}^{E,\infty}$ /kJ ·mol$^{-1}$ | $\Delta H_{\text{N-X}}^{\infty}$ |
|---|---|---|
| DEE + heptane | 1.6 [59] | |
| DPE + heptane | 0.84 [130] | |
| DBE + heptane | 0.53 [131] | |
| THF + heptane | 3.2 [132] | |
| THP + heptane | 2.2 [132] | |
| 1,4-dioxane + heptane | 8.4 [132] | |
| DMF + cyclohexane | 13.6 [70] | |
| DMA + cyclohexane | 12.8 [63] | |
| EtN + cyclohexane | 15.0 [71] | |
| BA + heptane | 5.9 [61] | |
| HxA + heptane | 5.7 [61] | |
| DEA + heptane | 3.04 [60] | |
| DPA + heptane | 1.96 [60] | |
| DBA + heptane | 1.54 [60] | |
| BA + DEE | 1.29 [74] | − 6.2 |
| BA + DPE | 1.83 [74] | − 4.9 |
| BA + DBE | 2.43 [74] | − 4.0 |
| BA + THF | 0.88 [74] | − 8.2 |
| BA + THP | 1.03 [74] | − 7.1 |
| BA + 1,4-dioxane | 2.03 [74] | − 12.3 |
| DEA + THF | − 0.014 [37] | − 6.2 |
| DEA + THP | − 0.043 [37] | − 5.3 |
| DEA + 1,4-dioxane | 2.25 [37] | − 9.2 |
| DPA + THF | 0.27 [37] | − 4.9 |
| DPA + THP | − 0.11 [37] | − 4.3 |
| DPA + 1,4-dioxane | 2.65 [37] | − 7.7 |
| DBA + DEE | 0.31 [129] | − 2.9 |
| DBA + DPE | 0.23 [129] | − 2.1 |
| DBA + DBE | 0.23 [129] | − 1.8 |
| DBA + THF | 0.71 [129] | − 4.0 |

Table 6 (continued)

| | | |
|---|---|---|
| DBA + THP | 0.47 [129] | − 3.3 |
| DBA + 1,4-dioxane | 3.47 [129] | − 6.5 |
| BA + DMF | 1.10 [34] | − 18.5 |
| HxA + DMF | 2.78 [34] | − 16.5 |
| DPA + DMF | 3.65 [34] | − 11.9 |
| DBA + DMF | 5.77 [34] | − 9.4 |
| BA + DMA | 0.68 [34] | − 18.0 |
| HxA + DMA | 1.93 [34] | − 16.6 |
| DPA + DMA | 2.76 [34] | − 12.0 |
| DBA + DMA | 4.59 [34] | − 9.7 |
| BA + EtN ($T$ = 303.15 K) | 2.65 [75] | − 18.2 |
| DEA + EtN | 2.92 [39] | − 15.1 |

[a]for symbols, see Table 1

TABLE 7

Results from the application of the Flory model to some alkyl-amine(1) + organic compound (2) systems at 298.15 K. Reduction parameters for pure compounds are reported in Table S1 (supplementary material)

| System[a] | $N$[b] | $X_{12}$ /J.cm$^{-3}$[c] | $dev(H_m^E)$ [d] | Ref. |
|---|---|---|---|---|
| BA + DEE | 20 | 13.53 | 0.064 | 74 |
| BA + DPE | 20 | 22.88 | 0.068 | 74 |
| BA + DBE | 23 | 26.72 | 0.066 | 74 |
| BA + 1,4-dioxane | 14 | 27.42 | 0.038 | 74 |
| DEA + 1,4-dioxane | 20 | 33.37 | 0.035 | 74 |
| DPA + 1,4-dioxane | 15 | 30.33 | 0.016 | 37 |
| DBA + 1,4-dioxane | 28 | 35.83 | 0.060 | 37 |
| BA + DMF | 15 | 21.53 | 0.087 | 129 |
| HxA + DMF | 14 | 29.31 | 0.025 | 34 |
| DPA + DMF | 14 | 33.52 | 0.007 | 34 |
| DBA + DMF | 14 | 41.82 | 0.030 | 34 |
| BA + DMA | 17 | 11.53 | 0.055 | 34 |
| HxA + DMA | 19 | 17.74 | 0.011 | 34 |
| DPA + DMA | 15 | 21.71 | 0.038 | 34 |
| DBA + DMF | 15 | 29.19 | 0.023 | 34 |
| DEA + EtN | 33 | 37.61 | 0.026 | 39 |

[a]for symbols, see Table 1; [b]number of data points; [c]interaction parameter in the Flory model; [d]equation (2)

TABLE 8

Results for solid-liquid equilibria of 1-alkylamine(1) + ethanenitrile(2) mixtures [102]

| Amine | $N$[a] | $\sigma_r(T)$ [b] | | $x_{1eu}$ [c] | | $T_{eu}$ [d] /K | |
|---|---|---|---|---|---|---|---|
| | | Ideal[e] | DQ[f] | Exp.[g] | DQ[f]. | Exp.[g] | DQ.[f] |
| 1-hexylamine | 28 | 0.037 | 0.003 | 0.043 | 0.046 | 228.2 | 229.2 |
| 1-octylamine | 30 | 0.058 | 0.008 | | | | |

[a]number of data points;[b] $\sigma_r(T) = \{\frac{1}{N}\sum\left[\frac{T_{exp}-T_{calc}}{T_{exp}}\right]^2\}^{1/2}$ [c]eutectic composition; [d]eutectic temperature; [e]result from the application of the ideal solubility model; [f]DISQUAC value calculated with interaction parameters from Table 5; [g] experimental value;

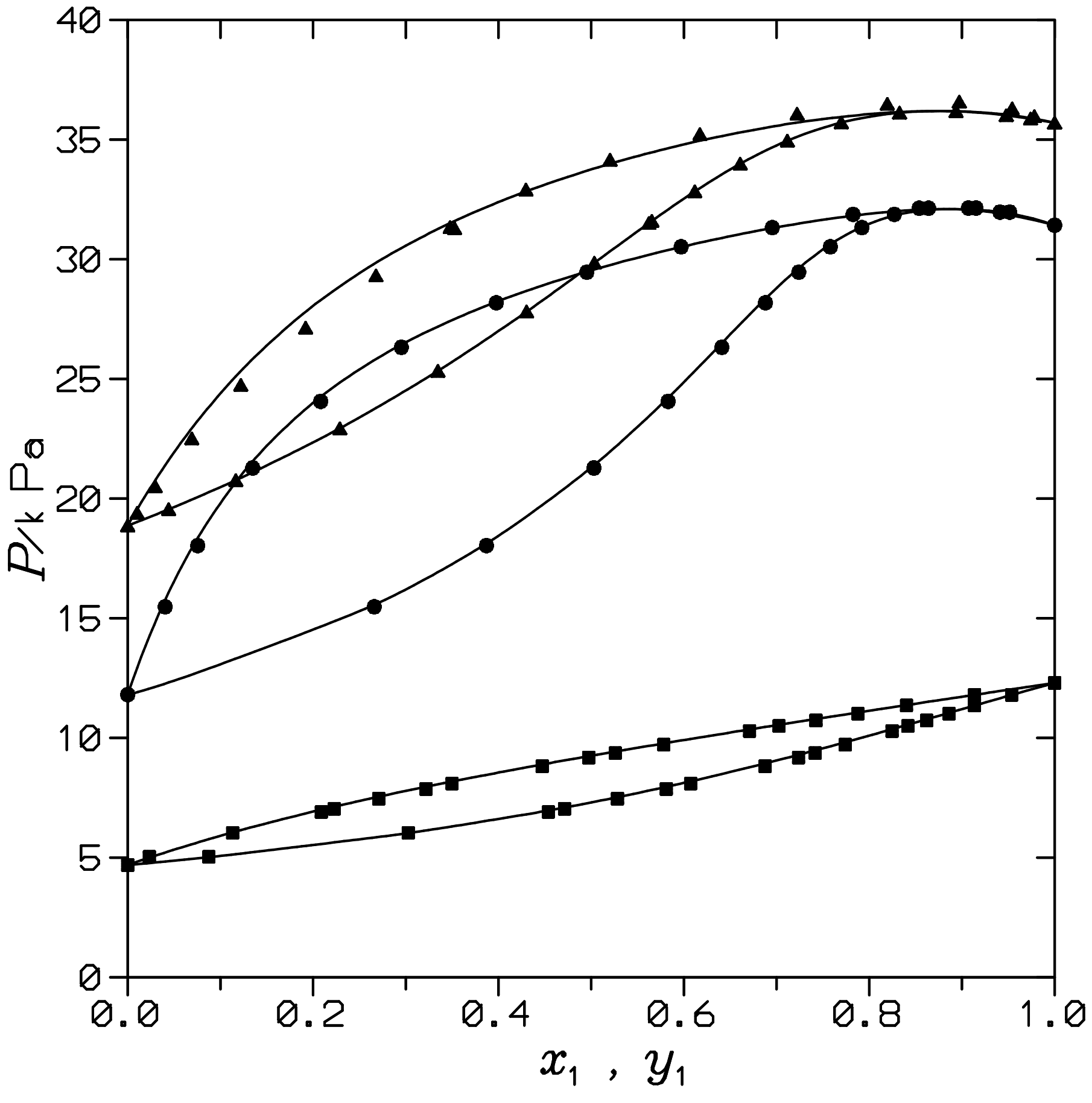

Figure 1. VLE of linear amine(1) + organic compound(2) mixtures. Points, experimental results: (■), 1-butylamine(1) + 1,4-dioxane(2) at 298.15 K [100]; (●), di-*n*-ethylamine(1) + ethanenitrile(2) at 298. K [101]; (▲), cyclohexylamine(1) + DMF(2) at 373.15 K [124]. Solid lines, DISQUAC calculations with interaction parameters from Table 5.

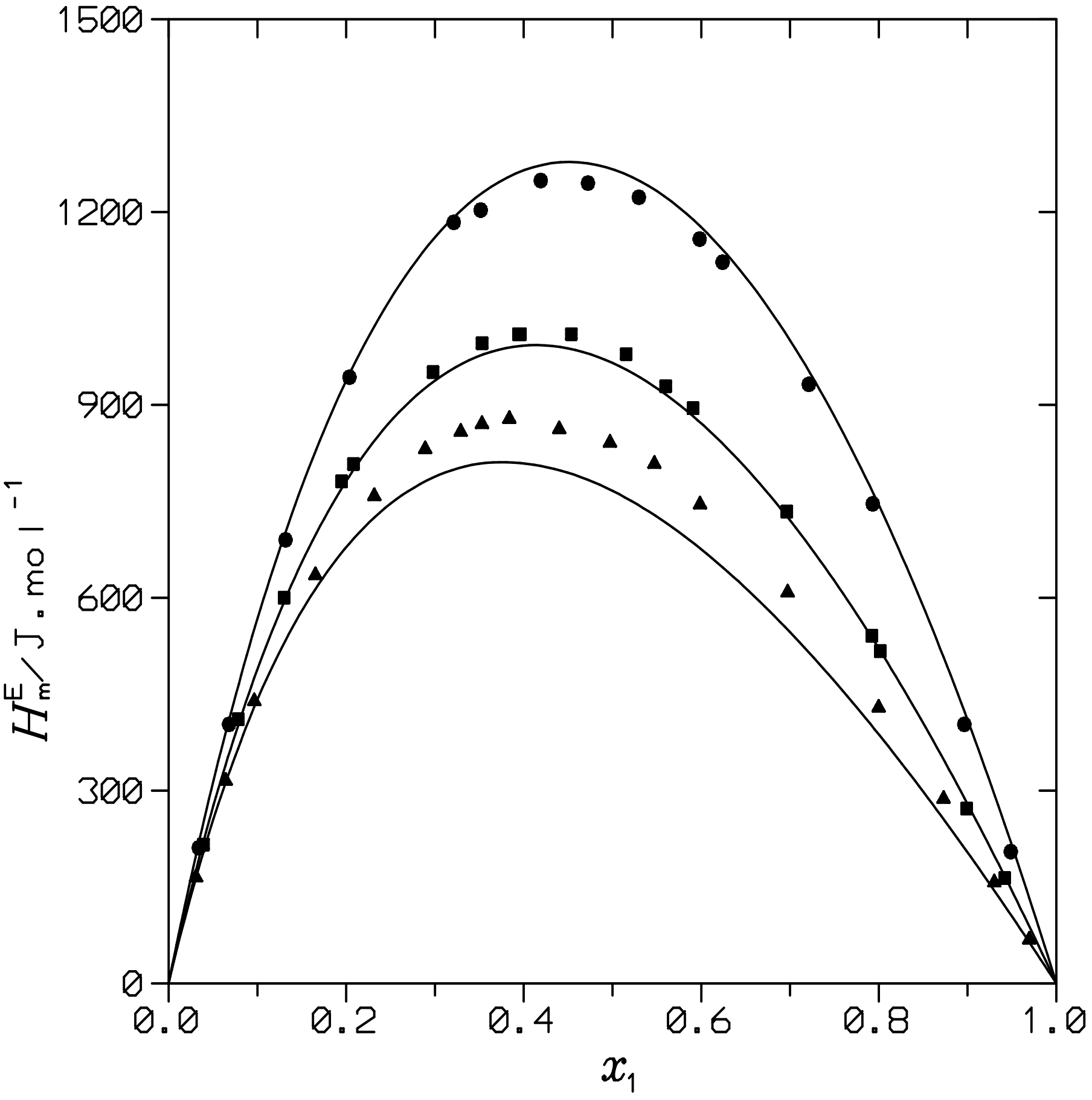

Figure 2. $H^{E}_{m}$ of 1-alkylamine(1) + cyclohexane(2) mixtures at 303.15 K. Points, experimental results [126]: (●), 1-propylamine; (■), 1-hexylamine; (▲), 1-decylamine. Solid lines, DISQUAC calculations with the interaction parameters from Table 5.

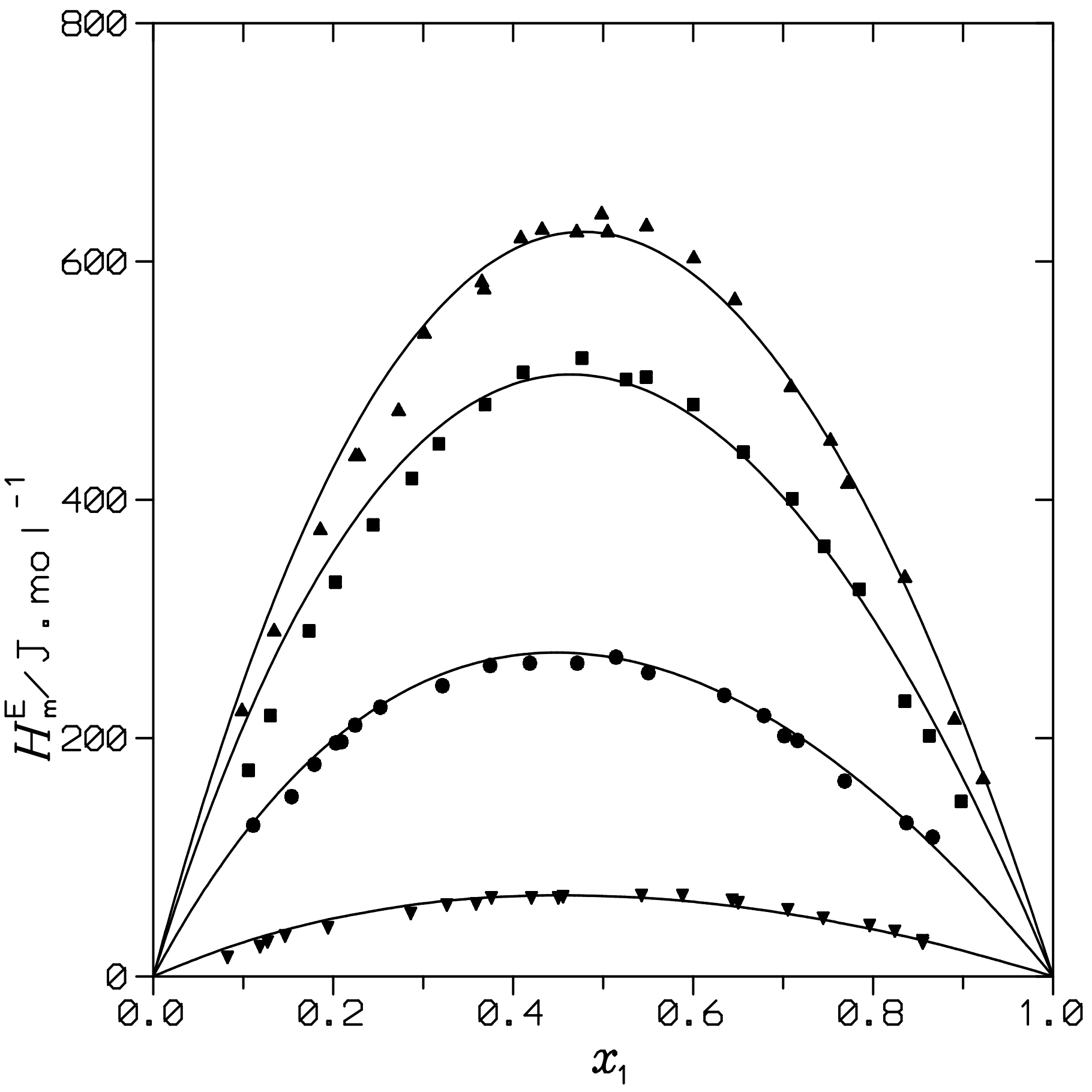

Figure 3. $H^{E}_{m}$ of linear amine(1) + linear monoether(2) mixtures at 298.15 K. Points, experimental results: 1-butylamine(1) + di-*n*-ethylether(2) (●), or + di-*n*-propylether(2) (■), or + di-*n*-butylether(2) (▲) [74]; di-*n*-butylamine(1) + di-*n*-propylether(2) (▼) [129]. Solid lines, DISQUAC calculations with the interaction parameters from Table 5.

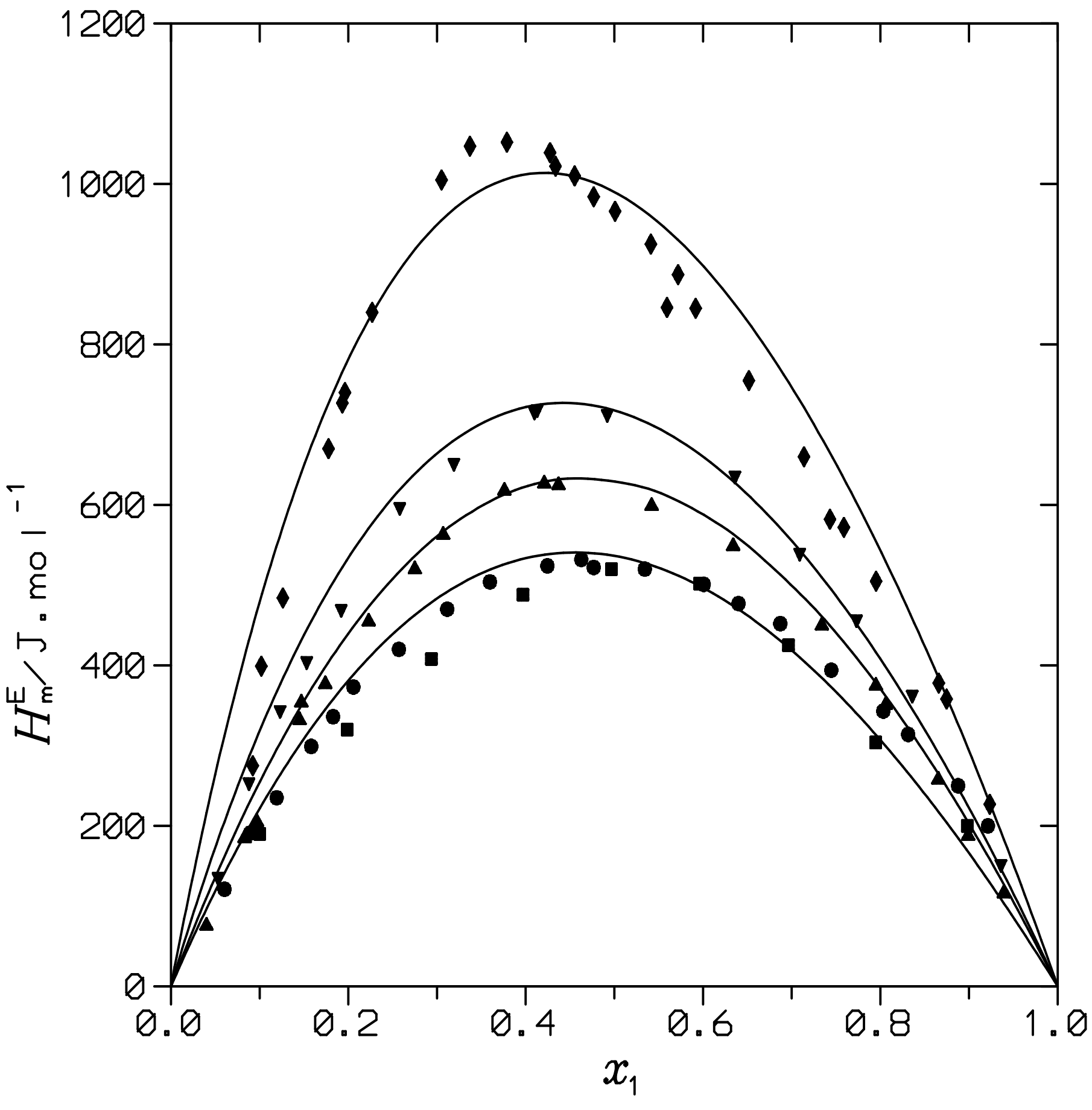

Figure 4. $H_m^E$ of linear amine(1) + 1,4-dioxane(2) mixtures at 298.15 K. Points, experimental results: (●) [74] and (■) [128], 1-butylamine; (▲) di-*n*-ethylamine [37]; (▼), di-*n*-propylamine [37]; (◆), di-*n*-butylamine [129]. Solid lines, DISQUAC calculations with the interaction parameters from Table 5.

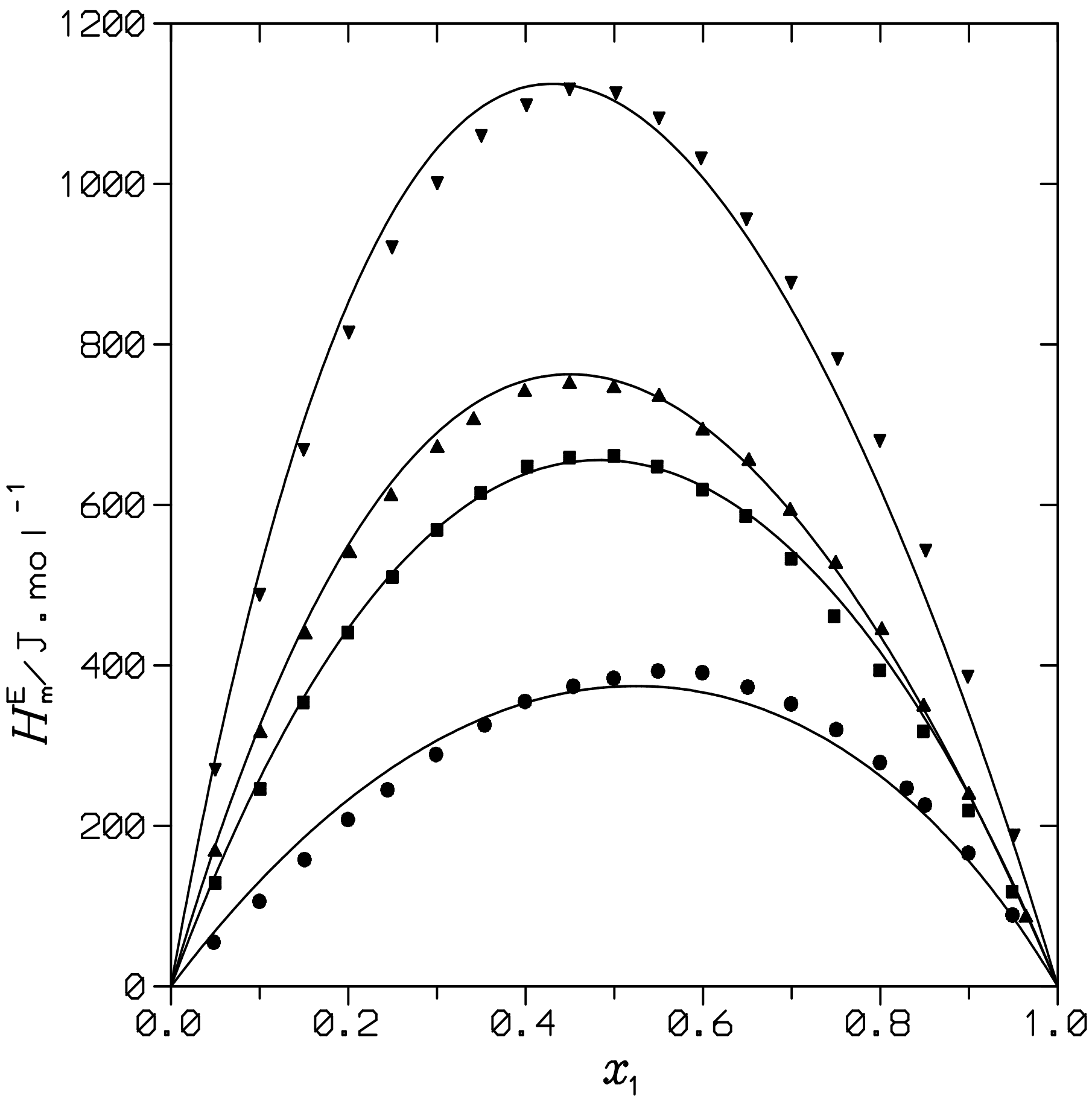

Figure 5. $H_m^E$ of linear amine(1) + DMF(2) mixtures at 298.15 K. Points, experimental results [34]: (●), 1-butylamine; (■), 1-hexylamine; (▲), di-*n*-propylamine; (◆), di-*n*-butylamine . Solid lines, DISQUAC calculations with the interaction parameters from Table 5.

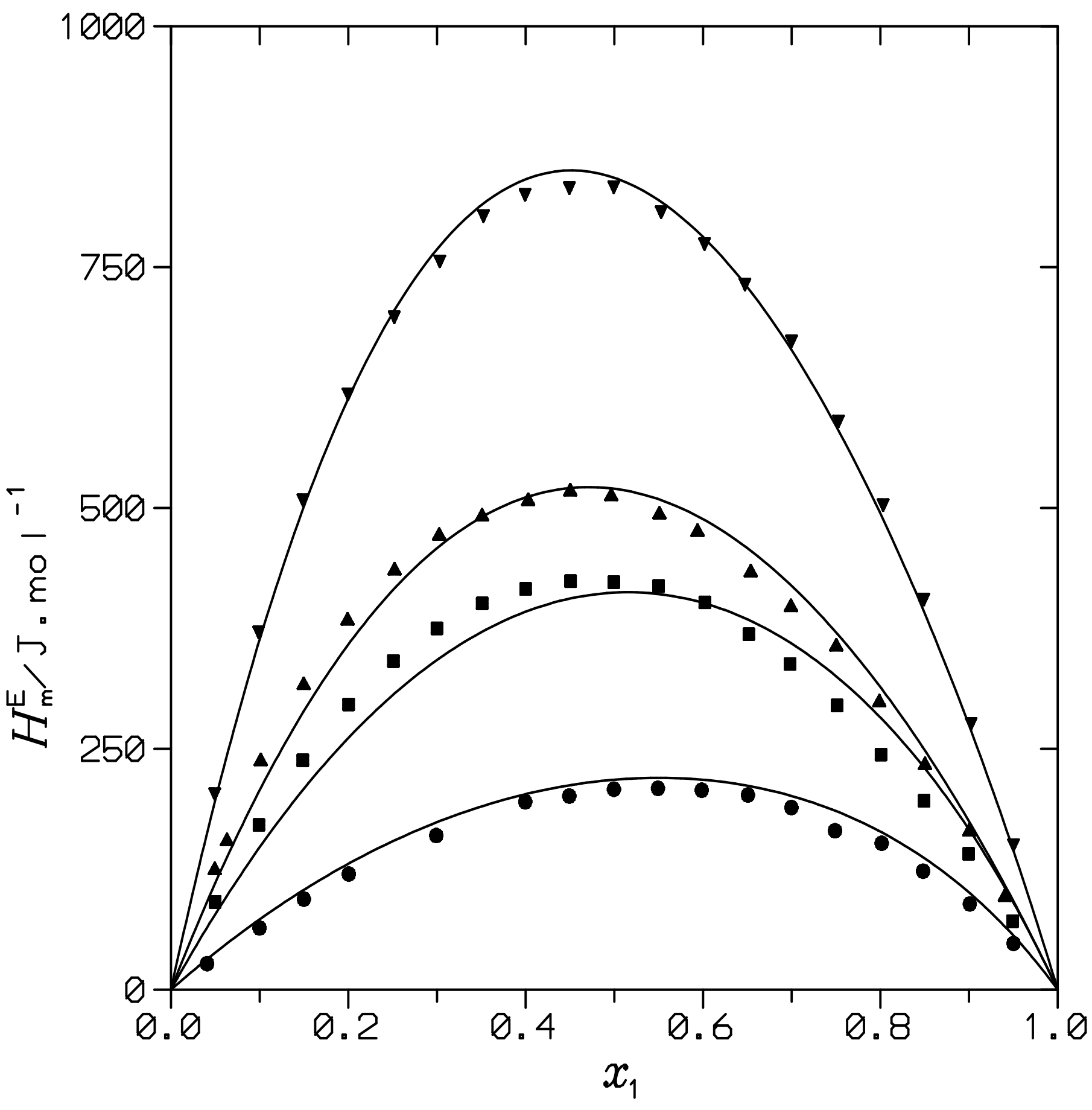

Figure 6. $H^{\mathrm{E}}_{\mathrm{m}}$ of linear amine(1) + DMA(2) mixtures at 298.15 K. Points, experimental results [34]: (●), 1-butylamine; (■), 1-hexylamine; (▲), di-*n*-propylamine; (◆), di-*n*-butylamine . Solid lines, DISQUAC calculations with the interaction parameters from Table 5.

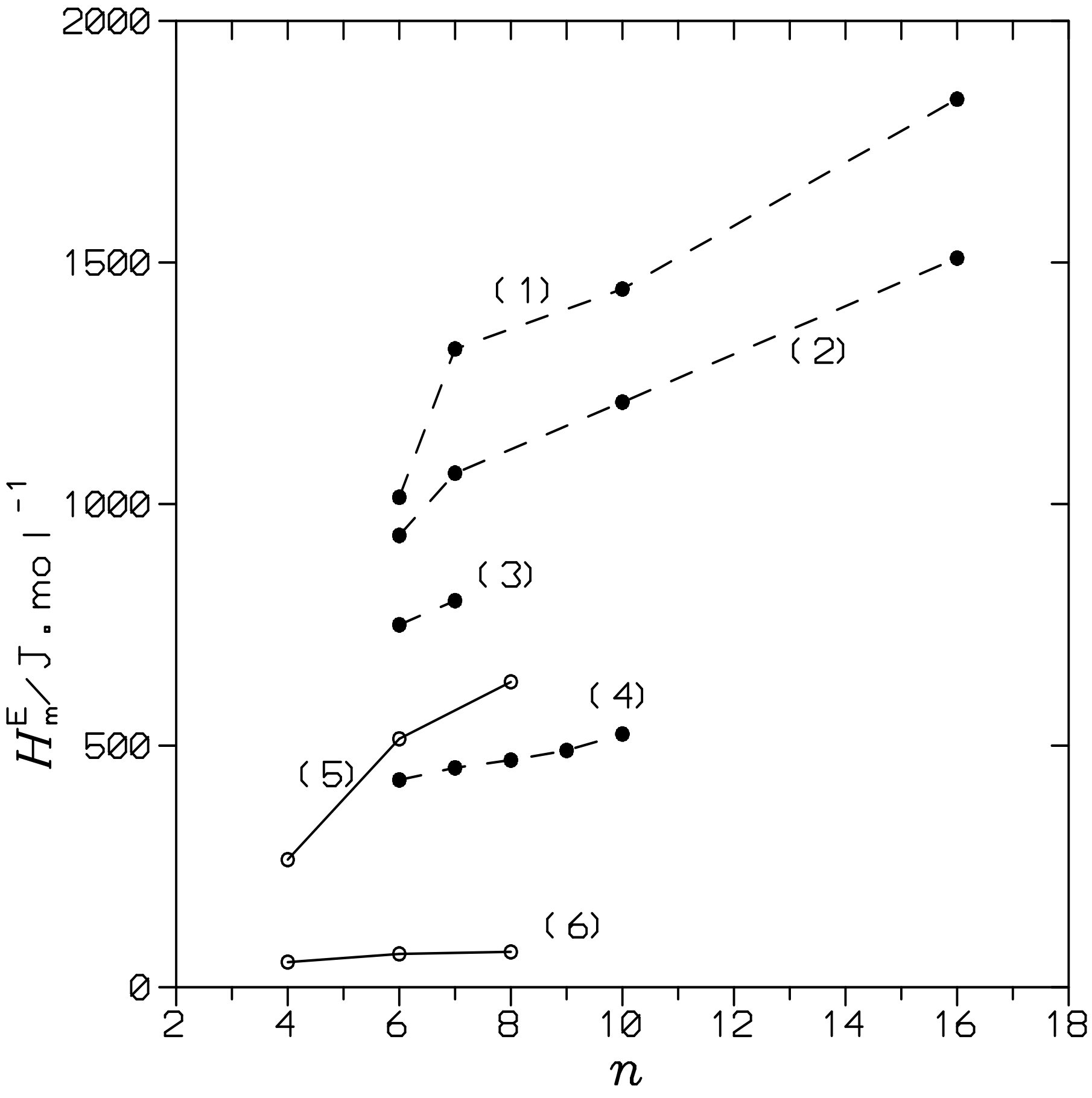

Figure 7 $H_m^E$ values at 298.15 K and equimolar composition for the mixtures: (1), 1-butylamine(1) + *n*-alkane(2) [133-135]; (2), 1-hexylamine(1) + *n*-alkane(2) [135,136]; (3), di-*n*-ethylamine(1) + *n*-alkane (2) [137] ($T$ = 293.15 K); (4), di-*n*-propylamine(1) + *n*-alkane(2) [138]; (5) 1-butylamine(1) + di-*n*-alkylether(2) [74]; (6) di-*n*-butylamine(1) + di-*n*-alkylether (2) [129]; *n* is the number of the C atoms in *n*-alkanes, or di-*n*-alkylethers. Lines are only for the aid of the eye

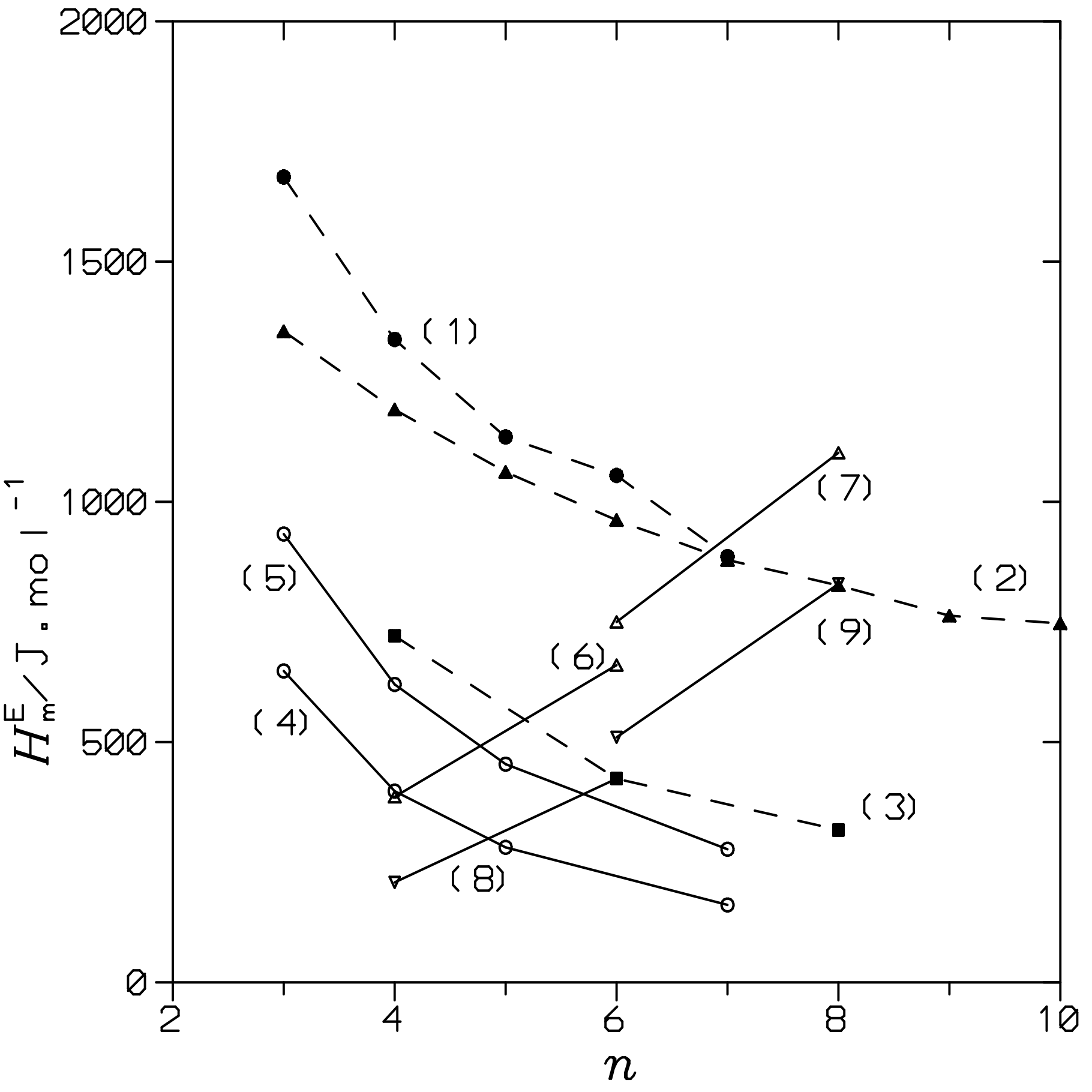

Figure 8 $H^E_m$ values at 298.15 K and equimolar composition for the mixtures: (1), 2-alkanone(1) + heptane(2) [139-141] ; (2), 1-alkylamine(1) + heptane(2) [61]; (3), di-*n*-alkylamine(1) + heptane(2) [60]; (4), 2-alkanone(1) + DPA(2) [12]; (5), 2-alkanone(1) + DBA(2) [12]; (6), 1-alkylamine(1) + DMF(2) [34]; (7), di-*n*-alkylamine(1) + DMF(2) [34]; (8), 1-alkylamine(1) + DMA(2) [34]; (9), di-*n*-alkylamine(1) + DMA(2) [34]; *n* is the number of the C atoms in alkylamines, or 2-alkanones. Lines are only for the aid of the eye.

# SUPPLEMENTARY MATERIAL

# THERMODYNAMICS OF AMINE MIXTURES. SYSTEMS FORMED BY ALKYL-AMINE AND ETHER, OR *N,N*-DIALKYLAMIDE, OR ETHANENITRILE

Juan Antonio González, [1] Fernando Hevia,[2] Isaías García de la Fuente,[1] José Carlos Cobos[1], Karine Ballerat-Busserolles[2], Yohann Coulier[2], Jean-Yves Coxam[2]

[1]G.E.T.E.F., Departamento de Física Aplicada, Facultad de Ciencias, Universidad de Valladolid. Paseo de Belén, 7, 47011 Valladolid, Spain.

[2]Institut de Chimie de Clermont Ferrand, Thermodynamique et Interactions Moléculaires, UMR CNRS 6296, University Clermont Auvergne, Aubière, France

*e-mail: jagl@termo.uva.es; Tel: +34-983-423757

TABLE S1

Properties[a] of pure compounds at $T$ = 298.15 K needed for the application of the Flory theory.

| Compund[b] | $V_i$ / cm$^3$ mol$^{-1}$ | $\alpha_p$ / $10^{-3}$ K$^{-1}$ | $\kappa_T$ / $10^{-12}$ Pa$^{-1}$ | $V_i^*$ / cm$^3$ mol$^{-1}$ | $P_i^*$ / J cm$^{-3}$ |
|---|---|---|---|---|---|
| BA | 99.88 [s1] | 1.311 [s2] | 1159 [s2] | 76.35 | 577 |
| HxA | 133.06 [s1] | 1.128 [s2] | 975 [s2] | 104.49 | 559.5 |
| DEA | 104.24 [s3] | 1.53 [s4] | 1471 [s4] | 77.38 | 562.7 |
| DPA | 138.07 [s3] | 1.31 [s5] | 1220 [s5] | 105.56 | 547.7 |
| DBA | 171.03 [s3] | 1.07 [s6] | 1020 [s6] | 135.54 | 498 |
| DEE | 104.74 [s7] | 1.654 [s7] | 1967 [s7] | 76.57 | 469.1 |
| DPE | 137.68 [s8] | 1.26 [s8] | 1440 [s8] | 106 | 440.7 |
| DBE | 170.45 [s9] | 1.134 [s9] | 1205.9 [s9] | 133.74 | 455.2 |
| THF | 81.76 [s10] | 1.2265[s10] | 962.3 [s10] | 63.26 | 634.7 |
| THP | 98.19 [s11] | 1.156 [s12] | 990 [s13] | 76.78 | 569.4 |
| 1,4-dioxane | 85.71 [s11] | 1.115 [s11] | 738 [s11] | 67.44 | 727.5 |
| DMF | 77.44 [s11] | 1.008 [s2] | 659.9 [s2] | 61.98 | 711.4 |
| DMA | 93.04 [s11] | 0.98 [s14] | 653.5 [s14] | 74.82 | 691.4 |
| EtN | 52.87 [s11] | 1.35 [s11] | 1070 [s11] | 40.20 | 650.8 |

[a] $V_i$, molar volume, $\alpha_p$, isobaric thermal expansion coefficient; $\kappa_T$, isothermal compressibility; $V_i^*$, reduction parameter for volume and $P_i^*$, reduction parameter for pressure in the Flory model; [b]for symbols, see Table 1.

References

[s1] E. Matteoli, L. Lepori, A. Spanedda, Fluid Phase Equilib. 212 (2003) 41-52.

[s2] F. Hevia, A. Cobos, J.A. González, I. García de la Fuente, L.F. Sanz, J. Chem. Eng. Data 61 (2016) 1468-1478.

[s3] E. Matteoli, P. Gianni, L. Lepori, Fluid Phase Equilib. 306 (2011) 234-241.

[s4] H. Funke, M. Wetzel, A. Heintz, Pure Appl. Chem. 61 (1989) 1429-1439.

[s5] T.M. Letcher, B.C. Bricknell, J. Chem. Eng. Data 41 (1996) 166-169.

[s6] N. Riesco, S. Villa, J.A. González, I. García de la Fuente, J.C. Cobos, Fluid Phase Equilib., 202 (2002) 345-358.

[s7] J. Canosa, A. Rodríguez, J. Tojo, Fluid Phase Equilib.156 (1999) 57-71.

[s8] R. Garriga, S. Martínez, P. Pérez, M. Gracia, Fluid Phase Equilib. 147 (1998) 195-206.

[s9] I. Mozo, I. García de la Fuente, J.A. González, J.C. Cobos, J. Mol. Liq. 129 (2006) 155-163.

[s10] M. Keller, S. Schnabel, A. Heintz, Fluid Phase Equilib. 110 (1995) 231-265.

[s11] J.A. Riddick, W.B. Bunger, T. K. Sakano, Organic Solvents. Physical Properties and Methods of Purification, Techniques of Chemistry, Volume II. John Wiley & Sons, A Weissberger, Ed., N.Y, (1986).

[s12] B. Giner, B. Olivares, I. Giner, G. Pera, C. Lafuente, J. Solution Chem. 36 (2007) 375-386.

[s13] I. Cibulka, L. Hnedkovský, T. Takagi, J. Chem. Eng. Data 42 (1997) 2-26.

[s14] F. Hevia, A. Cobos, J.A. González, I. García de la Fuente, V. Alonso, J. Solution. Chem., 46 (2017) 150-174.

[s15] R. Srivastava, B.D. Smith, J. Chem. Eng. Data 30 (1985) 308-313.

[s16] T.M. Letcher, P.U. Govender, U. Domanska, J. Chem. Eng. Data 44 (1999) 274-285.

[s17] U. Domanska, M. Marciniak, J. Chem. Thermodyn. 39 (2007) 247-253.

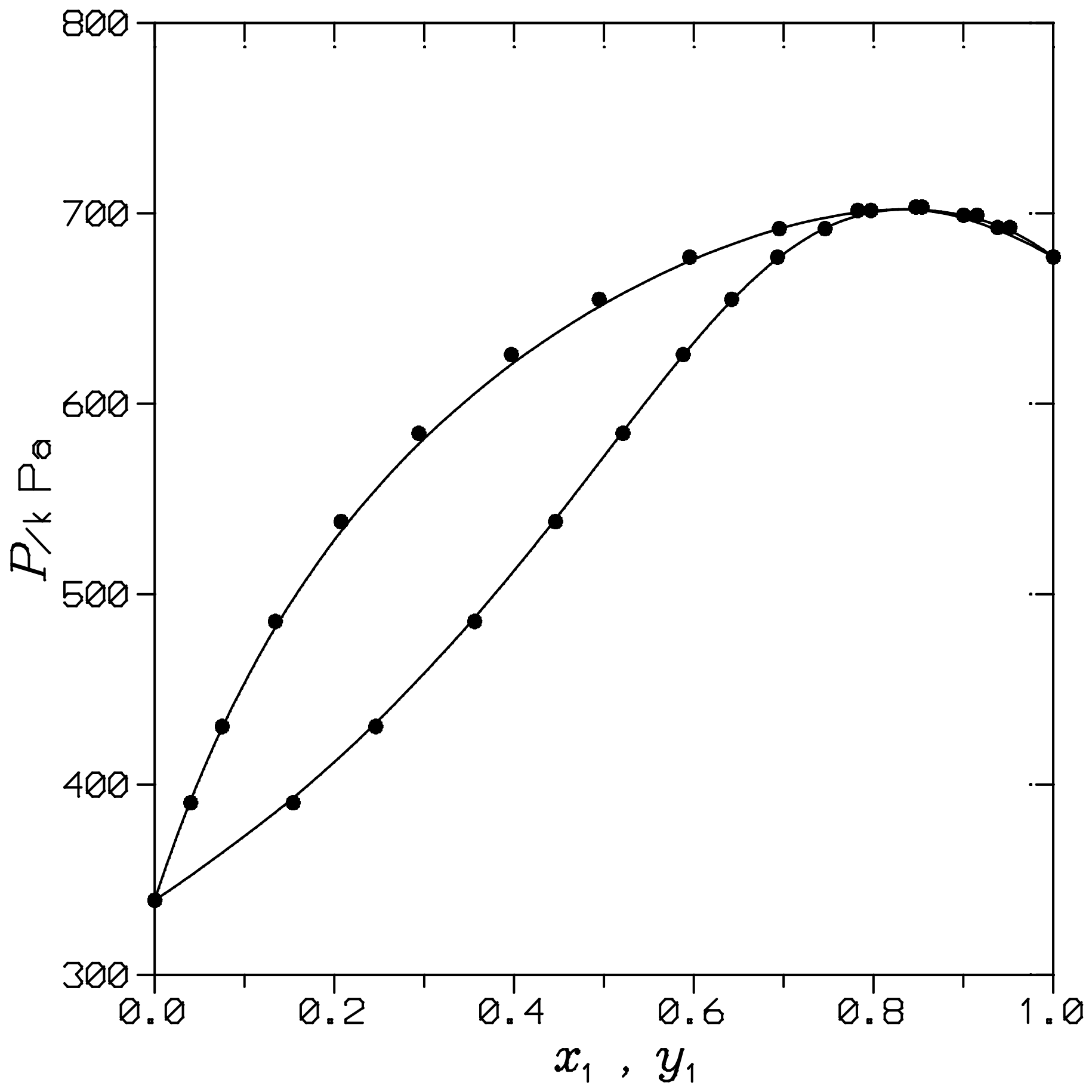

Figure S1 VLE for the di-n-ethylamine(1) + EtN(2) mixture at 398.33 K. Points experimental results [S15]. Lines, DISQUAC calculation with the interaction parameters from Table 5.

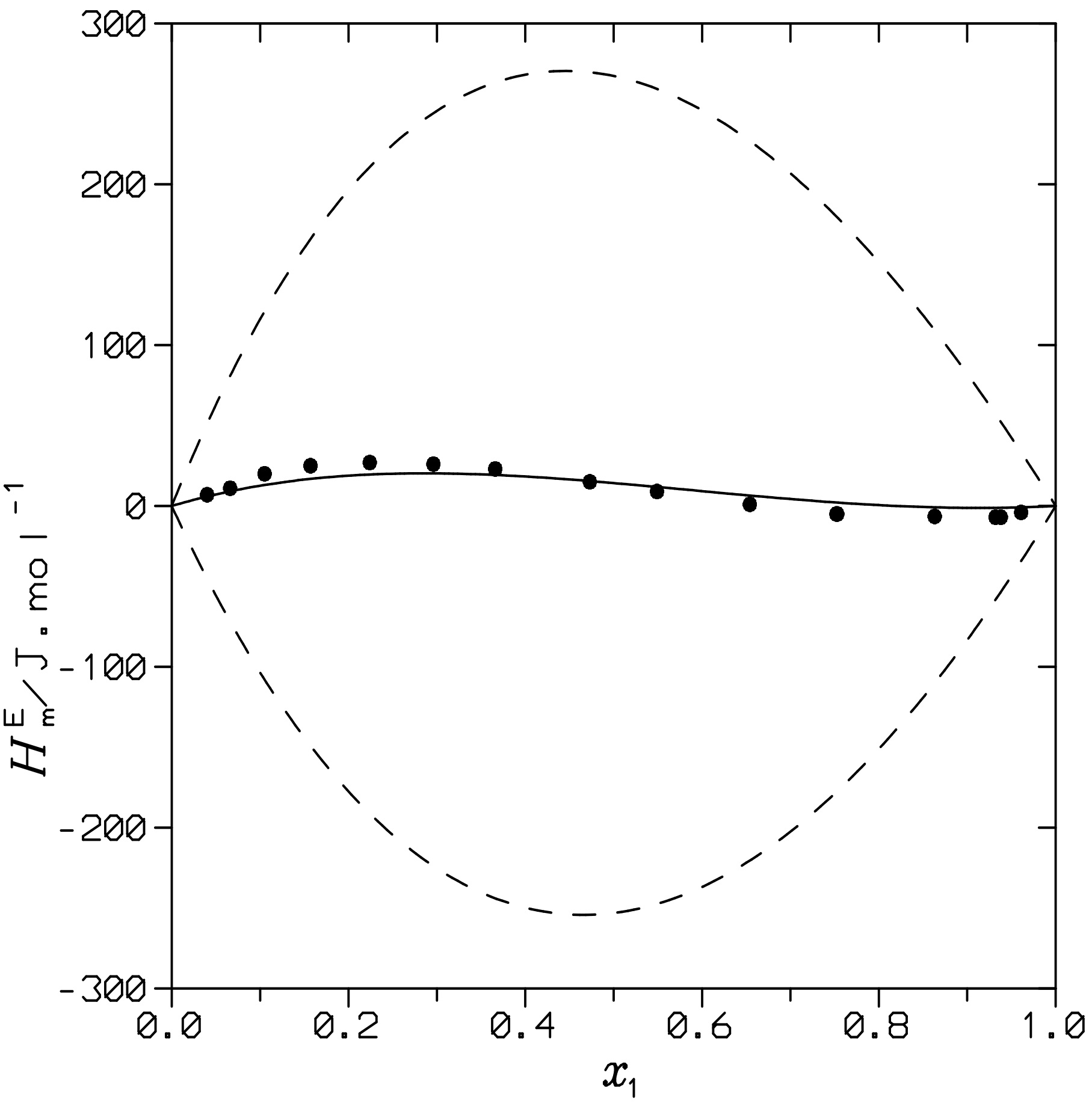

Figure S2 $H_m^E$ of di-*n*-propylamine(1) + THP(2) mixture at 298.15 K. Points, experimental results [S16]. Solid line, DISQUAC calculations with the interaction parameters from Table 5. Dashed lines, dispersive (upper curve) and quasichemical (lower curve) contributions to $H_m^E$.

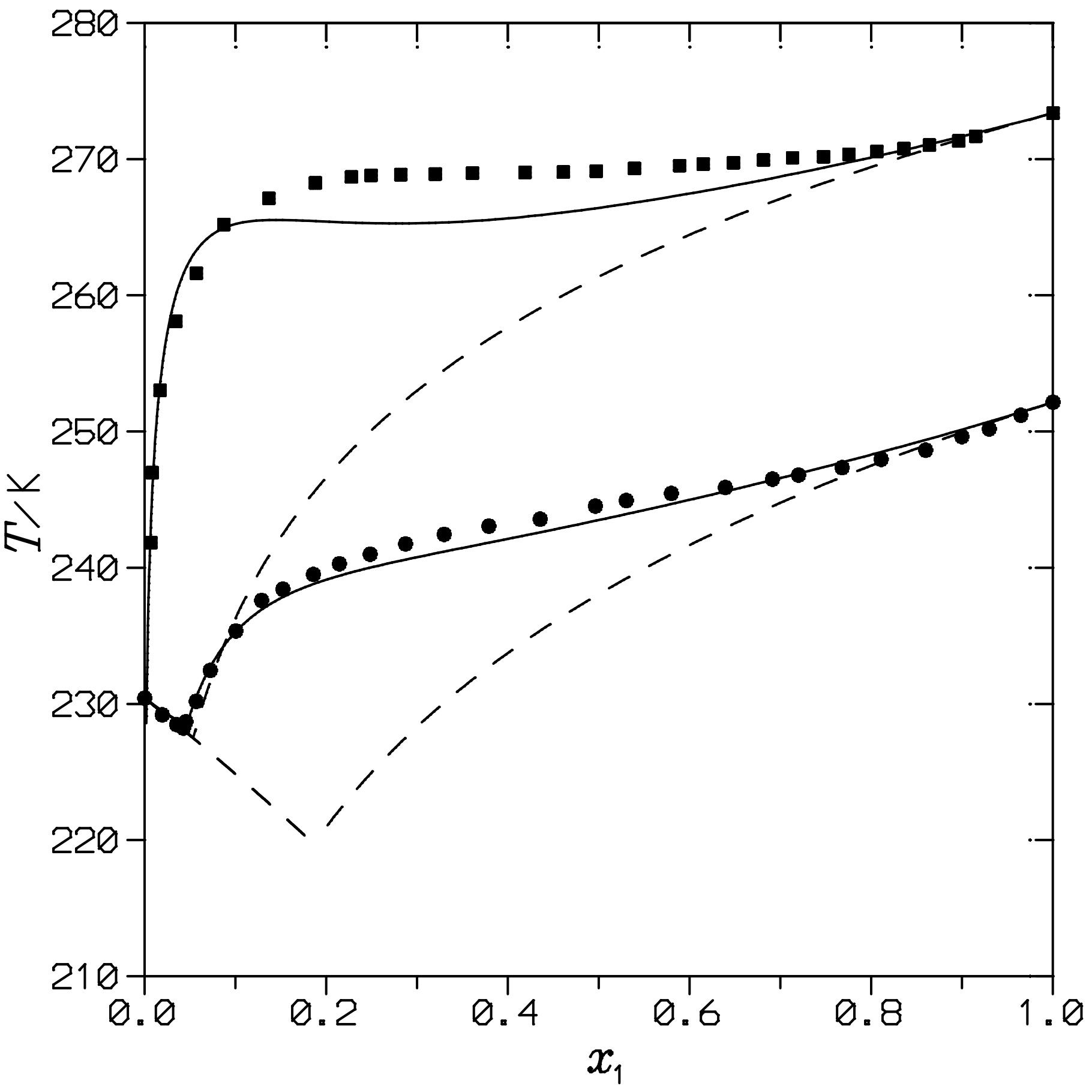

Figure S3 SLE for the 1-alkylamine(1) + EtN(2) mixtures. Points, experimental results [S17]: (●) 1-hexylamine, (■), 1-octylamine. Solid lines, DISQUAC calculations with the interaction parameters listed in Table 5. Dashed lines, results from the ideal solubility model.